\documentclass[12pt,preprint]{aastex}
\shorttitle{Galaxy activity and  Compact Group dynamics}
\shortauthors{Coziol, Brinks \& Bravo--Alfaro}

\begin{document}

\title{The relation between galaxy activity and the dynamics of
compact groups of galaxies}

\author{R. Coziol}
\affil{Departamento de Astronom\'{\i}a, Universidad de Guanajuato\\
Apartado postal 144, 36000 Guanajuato, Gto, Mexico}
\email{rcoziol@astro.ugto.mx}
\author{E. Brinks}
\affil{Instituto Nacional de Astrof\'{\i}sica, Optica y
Electr\'{o}nica, Apdo. Postal 51 \& 216, Puebla, Pue 72000,
Mexico} \email{ebrinks@inaoep.mx}
\author{H. Bravo--Alfaro}
\affil{Departamento de Astronom\'{\i}a, Universidad de Guanajuato\\
Apartado postal 144, 36000 Guanajuato, Gto, Mexico}
\email{hector@astro.ugto.mx}

\begin{abstract}
Using a sample of 91 galaxies distributed over 27 Compact Groups of Galaxies
(CGs), we define an index that allows us to quantify their level of activity,
be it AGN or star formation. By combining the mean activity index with the mean
morphological type of the galaxies in a group we are able to quantify the
evolutionary state of the groups. We find that they span a sequence in
evolution, which is correlated with the spatial configuration of the galaxies
making up a CG. We distinguish three main configuration Types, A, B and C.
Type~A CGs show predominantly low velocity dispersions and are rich in
late-type spirals that are active in terms of star formation or harbor an AGN.
Type~B groups have intermediate velocity dispersions and contain a large
fraction of interacting or merging galaxies. Type~C is formed by CGs with high
velocity dispersions, which are dominated by elliptical galaxies that show no
activity. We suggest that the level of evolution increases in the sense
A$\Rightarrow$B$\Rightarrow$C. Mapping the groups with different evolution
levels in a diagram of radius versus velocity dispersion does not reveal the
pattern expected based on the conventional fast merger model for CGs, which
predicts a direct relation between these two parameters. Instead, we observe a
trend that goes contrary to expectation: the level of evolution of a group
increases with velocity dispersion. This trend seems to be related to the
masses of the structures in which CGs are embedded. In general, the level of
evolution of a group increases with the mass of the structure. This suggests
either that galaxies evolve more rapidly in massive structures or that the
formation of CGs embedded in massive structures predated the formation of CGs
associated with lower mass systems. Our observations are consistent with the
formation of structures as predicted by the CDM model (or $\Lambda$CDM),
assuming the formation of galaxies is a biased process.

\end{abstract}


\section{Introduction}

Although it seems today an inescapable conclusion that the formation and
evolution of galaxies is influenced by their environment, the details of how
these processes occur in space and time are still largely unknown. One example
is that of compact groups of galaxies (CGs). As a result of our studies of the
activity in galaxies in CGs, we now have a better understanding of the
evolution of galaxies in these systems (Coziol et al. 1998a,b; Coziol, Iovino
\& de Carvalho 2000). Our observations showed emission-line galaxies to be
remarkably frequent, representing more than 50\% of the galaxies in Compact
Groups. Non-thermal activity, in the form of Seyfert~2s, LINERs and numerous
Low Luminosity AGNs (LLAGNs; see Coziol et al. 1998a for a definition of this
activity type in CGs), was found to constitute one important aspect of this
activity, whereas nuclear star formation, although mildly enhanced in some
groups, was noted to be generally declining. These observations were considered
to be consistent with the effects of tidal forces exerted on disk galaxies when
they fall into the potential well of a rich cluster or group of galaxies
(Coziol, Iovino \& de Carvalho 2000). Tidal stripping will remove gas from a
galaxy, reducing star formation in the disk, whereas tidal triggering will
start a short burst of star formation in the nuclear region and fuel an AGN
(Merritt 1983, 1984; Byrd \& Valtonen 1990; Henrikson \& Byrd 1996; Fujita
1998).

The above processes may also produce the density-morphology-activity relation
observed in CGs (Coziol et al. 1998a). By losing their gas and forming new
stars near their nucleus, increasing their bulge, the morphology of spiral
galaxies falling into groups is transformed to an earlier type. Assuming that
groups are continuously replenished by spiral galaxies from the field
(Governato, Tozzi \& Cavaliere 1996; Coziol, Iovino \& de Carvalho 2000), the
cores of the groups are naturally expected to be populated by quiescent
galaxies and AGNs (LLAGNs included), all having an early-type morphology, and
their periphery to be richer in late-type star forming galaxies.

What is missing in the above description is a connection between galaxy
evolution and the physical processes responsible for the formation and
evolution of CGs. Our first interpretation of these systems, based on
galaxy-galaxy interaction models, suggested they could not survive mergers over
a long period of time (Barnes~ 1989), which seems in conflict with the high
number of CGs observed today (for a different point of view, see Aceves \&
Vel\'azquez 2002). It was then realized that this paradox may be explained, in
part, by the fairly simplistic assumptions made about the nature of these
systems. For example, if CGs are associated with larger and dynamically more
complex structures, as many redshift surveys suggest (Rood \& Strubble 1994;
Ramella et al. 1994; Garcia 1995; Ribeiro et al. 1998; Barton, de Carvalho \&
Geller 1998), their formation and evolution must also be more complicated than
previously thought.

In this study, we examine further the question of the formation and evolution
of CGs, by extending our analysis of the activity to a larger sample of Hickson
Compact Groups of galaxies (HCGs: Hickson 1982).

\section{Observation}

Long-slit spectra of 65 galaxies in 19 HCGs were obtained with the 2.1m
telescope located at San Pedro M\'artir, in Baja California, during one run of
4 nights in April and another of 5 nights in October 2002. A Boller \& Chivens
spectrograph was used in conjunction with a SITE $1024 \times 1024$ CCD,
yielding a plate scale of 1.05 arcsec per pixel. During the observations, the
slit width was kept open to 240 micron, which corresponds to 3.1 arcsec on the
sky. This aperture was chosen to be slightly larger than the effective seeing
of 2 arcsec (which includes dome seeing and instrument response). We used a 300
l/mm grating, blazed at 5000 \AA, which, with the slit width adopted, yielded
an average resolution of 8 \AA\ over a 4096 \AA\ range.

Table~1 shows an excerpt of our observation logbook. Column~1 gives the date of
observation, whereas column~2 identifies the object (HCG numbers and letters,
as given by Hickson (1982) and as archived in CatalogVII/213 at the ``Centre de
Donn\'ees astronomiques de Strasbourg'' (CDS)). The positions of the galaxies
observed in column~3 and 4 are those given in the CDS catalog, precessed by us
for the year 2000. The heliocentric velocities, column~5, and the morphologies
of the observed galaxies, columns 6 and 7, were also taken from the same
source. Only galaxies pertaining to the groups (based on their redshifts) were
observed. Note that, due to observational constraints, we did not succeed in
observing all galaxies in each group, as originally planned. Our sample is
therefore somewhat incomplete and biased towards the most luminous galaxies in
each group. The effect this bias might have on our analysis is discussed at
length in Section~4.3.

Column~8 of Table~1 gives the number of exposures and their durations. In
general, we took three spectra of 900s each. The long slit (covering 322 arcsec
on the sky) was always centered on the most luminous part of the galaxies and
usually aligned in the E--W direction, to minimize errors introduced by
guiding. In very few cases, the slit was rotated to a different angle to put
more than one galaxy in the slit. These cases are identified by a plus sign in
front of the HCG number of the galaxy. The effective airmass is given in the
last column. Most of the observations were performed under low airmass
condition, minimizing differential diffraction.

Before the first exposure, and after each of the subsequent ones, a He-Ar lamp
was observed for wavelength calibration. During each night between two and
three standard stars were observed to flux calibrate the spectra. During our
observations we noted the frequent passage of light clouds during the night or
the presence of cirrus at the beginning or end of the night, which suggests
that the conditions were not photometric. Absolute flux calibration, however,
is not critical for our analysis.

\section{Results}

\subsection{Reduction and template subtraction}

Standard reduction techniques were followed using IRAF\footnote{IRAF, the Image
Reduction and Analysis Facility, a general purpose software system for the
reduction and analysis of astronomical data, is written and supported by the
IRAF programming group at the National Optical Astronomy Observatories (NOAO)
in Tucson, Arizona (http://iraf.noao.edu/).}. After bias subtraction, the
spectra were divided by a normalized dome flat. The sky was subtracted before
reducing the spectra to one dimension. After calibrating the one-dimensional
spectra in wavelength, the spectra were flux calibrated. The uncertainty in the
flux calibration is estimated to be 10\%. For each galaxy, an average spectrum
was obtained by combining the complete set of one-dimensional spectra (usually
three) available for that object.

The apertures used for the  reduction to one dimensional spectra contain 99\%
of the light from the most luminous part of the galaxies (the nucleus). The
equivalent linear size for the spatial apertures (Ap.) used are listed in
Table~2. It is interesting to note how compact the light distribution is in all
these galaxies. For the emission-line galaxies in particular, such compactness
suggests that the ionized gas covers regions extending out to only a few kpc
around the nucleus (the only exception being HCG 56a). A comparison with the
H$\alpha$ imaging study performed by Severgnini et al. (1999), which lists the
isophotal fluxes and luminosities of H$\alpha+$NII in an area 1 $\sigma$ above
the background for a sample of 73 HCGs, suggests that our spectroscopic survey
may have missed at most a few regions of low star formation activity in the
disk of some galaxies. Such type of activity does not seem to be predominant in
CGs.

Some 75\% of the emission-line galaxies in the groups show strong Balmer
absorption lines which needs to be corrected for before measuring line ratios
in these galaxies. To accomplish this we subtracted four different templates,
using the following quiescent galaxies from our sample: HCG~10b, 15c, 30a and
58d. Since we are interested only in classifying the galaxies, we did not try
to fit the templates to the entire candidate galaxy spectrum. Instead, we
concentrated on obtaining a good fit for the Balmer lines. The method followed
is straightforward and easy to apply.  First, all spectra are corrected to zero
redshift. Then a region 400 \AA\ wide around H$\alpha$ and 1500 \AA\ wide
around H$\beta$ is selected, to fit the continuum on each side of the lines.
Note that a larger wavelength range is needed for H$\beta$ in order to bridge
the strong Mg absorption bands visible in this region of the spectrum. Each
spectrum is then normalized, dividing it by the fitted continuum. The
templates, which were normalized the same way, are then simply subtracted from
the candidate galaxies. One example of a template subtraction is shown in
Figure~1.

The emission lines are measured in the template--subtracted spectrum by
integrating the flux under the lines. For H$\alpha$, the IRAF deblend routine
in SPLOT was used to correct for any contribution from [NII]. We only
determined line fluxes in those cases where after template subtraction the line
flux is larger than the rms in the residual (at full resolution). The H$\alpha$
flux is usually higher than 3$\sigma$ and it is higher than or equal to
2$\sigma$ for H$\beta$. In 10 cases only one of the lines (mostly H$\alpha$)
was detected after template subtraction. For the line ratios, we calculated the
mean of four values measured after subtracting one by one each of the
individual four templates. We adopted the dispersion of the mean as the
uncertainty. The two most important line ratios [NII]/H$\alpha$ and
[OIII]/H$\beta$ are given in Table~2. A plus sign in front of the HCG number
identifies the few emission-line galaxies for which a template subtraction was
not deemed necessary, as visual inspection of the spectra did not show any
signs of underlying absorption. Note that we did not apply any correction for
dust extinction, since these two line ratios are known to be insensitive to
this (Veilleux \& Osterbrock 1987). In Table~2, we also give the H$\alpha$
luminosity and equivalent width (EW) of the H$\alpha+$[NII] lines as measured
in the template subtracted spectra. The uncertainty for the luminosity is of
the order of 10\%. Luminosities with uncertainties larger than 20\% are
identified by a colon (uncertainties are always smaller than 30\%). The
uncertainty on the EW, which is mainly due to the template subtraction, is of
the order of 1 \AA.

\subsection{Activity classification}

Figure~2 shows the final spectrum for the galaxy HCG~58a, which is the most
active galaxy in our sample. As can be seen in the enlarged region, the
presence of weak, broad components around H$\alpha$ (also visible around
H$\beta$) suggests this is a Seyfert~1.8 (Osterbrock 1981). We emphasize that
HCG~58a is the first Seyfert~1 galaxy (although of an intermediate type) that
we have found thus far, after classifying 118 galaxies in 34 HCGs. This result
confirms the rarity of luminous AGNs observed in CGs (Coziol et al. 1998a;
Coziol, Iovino \& de Carvalho 2000).

The standard diagnostic diagram (Veilleux \& Osterbrock, 1987) for the
emission-line galaxies in our sample is shown in Figure~3 (the Seyfert~1.8 is
not included). It suggests that the LLAGNs are either LINER or Seyfert~2. The
spectra of these galaxies are shown in Figure~4a and 4b. For comparison, we
also show, in Figure~5a, the four LLAGN candidates that we were not able to
classify. The similarities in the spectra are obvious. Note in particular the
weakness of the H$\alpha$ line in these galaxies. This weakness is what prompts
us to classify them as possible AGNs. In contrast, we show in Figure~5b the
spectra of those galaxies which show weak star formation activity (Star Forming
Galaxies or SFGs). We see that the H$\alpha$ line is stronger than the two
nitrogen lines in these galaxies, which is a clear indication for star
formation. Another feature in common with luminous SFGs is their late-type
morphology (see the morphological classification in Table~2, column~3). The two
exceptions are HCG~56e and 79b.

The activity type adopted for each galaxy is reported in Table~2 (column~7).
They are identified as: quiescent galaxy (No em.), star forming galaxies (SFG),
LINER, Seyfert~2 (Sy2), Seyfert~1 (Sy1) and LLAGNs (dSy2 and dLINER ). A
question mark after the activity type indicates some uncertainty due to there
being only one emission line complex (either around H$\alpha$ or H$\beta$)
detected after template subtraction. This was the case for 10 galaxies which we
tentatively classified as follows: four LLAGN candidates and six weak SFGs.

Due to the difference in morphology between the templates used and the
emission-line galaxies, we expect our line ratios to be somewhat
underestimated. However, we note that to change the nature of the LLAGN from
LINER to star forming galaxies in the diagnostic diagram, we would need to have
underestimated the H$\alpha$ fluxes by a factor of more than 20, which seems
unreasonably large (see Miller \& Owen 2002, for a similar result in clusters
of galaxies).

In Figure~6, we compare the H$\alpha$ luminosity of the emission-line galaxies
with the equivalent width of the H$\alpha+$[NII] lines. The horizontal lines
indicate the mean H$\alpha$ luminosity (dot-dash) and lower limit as observed
in Nuclear Starburst Galaxies (Contini, Consid\`ere \& Davoust 1998). The
vertical (dashed) is the dividing line between galaxies with weak and active
star formation (Kennicutt \& Kent 1983). It can be seen that very few SFGs in
our sample qualify as a starburst or an actively star forming galaxy. In
Figure~6, one can see that what seems to distinguish most LINERS and LLAGNs
from SFGs are their slightly lower luminosity and distinctly smaller EW. The
difference in EW, in particular, is important, since it suggests the continuum
in these galaxies has a different origin (by definition, the EW is the ratio of
the flux in the emission line to the continuum). This difference could be a
sign of a weak AGN or of a significantly different ionizing stellar population.

Although we believe our classification is fair, we remind the reader that there
is an ongoing discussion about the AGN nature of LLAGNs and LINERs (see the
reviews by Lawrence 1999, and Ho 1999). Obviously, more observations will be
needed before we can understand what the role of star formation and AGN
activity is in these galaxies (see for example Turner et al. 2001).

In Table~3, we compare the distribution of type of activity in the present
sample with that obtained in our previous analyses (Coziol et al. 1998a;
Coziol, Iovino \& de Carvalho 2000). For comparison, the luminous LINER and
Seyfert galaxies were included in the same AGN bin. The distribution of
activity type in our new sample is fully consistent with what we observed
before.

\subsection{Activity index}

In order to quantify the level of activity in the groups, we define a new
spectroscopic index for the galaxies. Since we want this index to be sensitive
to the different stellar populations, and not only the ionized gas, we use the
equivalent width as measured in a window centered on H$\alpha$ (Copetti,
Pastoriza \& Dottori 1986; Coziol 1996). The size of the window was chosen by
considering the typical EWs of H$\alpha$ lines (in absorption) in the spectra
of A stars, which are known to dominate the stellar population after most of
the OB stars have disappeared (like in post-starburst or gas stripped galaxies;
see Rose 1985; Leonardi \& Rose 1996; Barbaro \& Poggianti 1997), and which are
responsible for the strong absorption features observed in LLAGNs (Coziol et
al. 1998a). We define the activity index, EW$_{\rm act}$, as the equivalent
width measured in the rest frame of the galaxy, between 6500 \AA\ and 6600 \AA\
in the original spectra (i.e., without template subtraction). Defined in this
way the activity index covers all the possible active states of a galaxy:
active (star formation and/or AGN), decreasing star formation (post-starburst
or gas stripping) and quiescent (intermediate and old stellar populations).
This way of defining the activity index has the advantage that it can be easily
measured in all the galaxies. The value of the activity index as measured in
each galaxy in our sample is given in the last column of Table~2. The
uncertainty in this value is of the order of $\pm0.2$\AA. It was estimated
using the mean of the EW of all the structures (real or spurious) encountered
in the continuum around the window in the four template galaxies. Note that,
because of their redshifts, the O$_2$ atmospheric band, at 6870 \AA, fell right
within the window of the galaxies in HCG~8, preventing us to measure an
EW$_{act}$.

In order to increase our sample of galaxies in CGs, we have measured the
activity index in those galaxies that were observed in our first
spectroscopical analysis of HCGs (Coziol et al. 1998a). Only four galaxies from
this previous article were duplications in the present study: HCG 40a, b, e and
HCG 88b. The activity indices found for these galaxies based on the 4m spectra
are 0.9, 3.6, -3.0 and -5.0 \AA\ respectively. This compares very well with the
activity indices measured in our new spectra: 1.0, 3.1, -3.2 and -6.1 \AA\
respectively.

Not considering two groups that we re-observed in the present study, and five
groups where only one galaxy of each group was observed, we can add 10 CGs to
our sample. The results are presented in Table~4, together with the
morphological type and activity classification for each galaxy.

\section{Analysis}

\subsection{Utility and robustness of the activity index}

Ideally, by using the mean of the activity index we should be able to
discriminate between active and non-active CGs. With active CGs we mean those
groups which harbor star forming galaxies and/or AGNs. To verify this, we
compare, in Figure~7a, the activity index as measured in galaxies with
different activity types. Since the range of the index is large, we take the
logarithm and discriminate between absorption (EW$_{act}>0$) and emission
(EW$_{act}<0$) using different symbols. In general, SFGs and luminous AGNs
(Sy1, Sy2 and LINER) have EW$_{act}<-6$ (or $\log |{\rm EW}_{act}|)> 0.8$).
This property allows us to separate them from quiescent galaxies and LLAGNs. We
therefore adopt a mean of EW$_{act}<-6$ to separate active from inactive
groups. Note that for a mean EW$_{act}> -6$, a negative mean index would
indicate some activity, but at a significantly lower level than for active
groups.

In Coziol et al. (1998a), the type of activity presented by the different
galaxies was found to be correlated with their morphology. To verify if this
phenomenon is also observed in the present sample, we compare the activity
index with the morphological type of the galaxies in Figure~7b. In general, we
find the SFGs in late-type (T $> 3$) galaxies and the quiescent galaxies in
early-type (T $< 1$). The exceptions are HCG~56d (S0) and 40d (SBa) for the
SFGs and HCG~30a (SBc) for the quiescent galaxies. We also note that AGNs and
LLAGNs are usually found in intermediate and early type galaxies.

In Figure~8, we repeat our analysis for the galaxies in the complementary
sample studied by Coziol et al. (1998). We distinguish the same trends as
observed in our new sample. This result suggests that the activity index is
independent of the type of telescope and reduction aperture used. This is most
readily attributed to the compact nature of the light and ionized gas
distribution in these galaxies (see our comment in Section~3.1).

\subsection{Definition of evolutionary levels for CGs}

As noted previously in Coziol et al. (1998a), the relation between morphology
and activity as found in CGs is not specific to these systems, but is general
to all galaxies independent of their environment (see e.g., Kennicutt 1992).
Taken as a group, however, it was shown in the same study that the
morphology-activity relation was correlated with the density of the groups.
This suggests that we should combine the mean activity index with the mean
morphological type of galaxies in a CG in order to quantify their level of
evolution. If activity is a consequence of galaxy interactions, and if
interactions lead to the formation of early--type systems, we would thus expect
''evolved'' groups to be inactive and rich in early-type galaxies, whereas
''unevolved'' groups would be active and rich in late-type galaxies. Another
possibility is for groups to be unusually active (or inactive) considering the
morphologies of their members. In the interaction scenario, such unusually
active groups may constitute a genuine intermediate evolutionary level.

The mean morphological types and mean activity indices of the 27 CGs in our
joined samples are reported in Table~5 (~columns~10 and 11, respectively). Note
that since only one member galaxy was observed for HCG~31 this group was
omitted from the rest of our analysis. In this table we have also collected
other useful information on the groups as found in the literature: the systemic
redshifts (column~2), as listed by Hickson in CatalogVII/213 (at the CDS); a
flag indicating diffuse X ray emission (column~3), as determined by Ponman et
al. (1996); the number of galaxies for which we have a spectral classification
(N$_S$; column~4); the number of real members (N$_H$; column~5), as determined
by Hickson (1982); the number of possible members (N$_R$; column~6), as
determined by Ribeiro et al. (1998); the de-projected velocity dispersions
(column~7), projected radii (column~8) and adopted masses (column~9) for the
groups, as listed by Hickson.

\subsection{Effect of incompleteness on our analysis}

As mentioned in Section~2, our information on the activity of the groups is not
complete due to observational constraints. To establish how this incompleteness
might affect our analysis, we have assigned an activity type to the missing
galaxies based on the trends between morphology and activity observed in
Figure~7 and 8 and used this prediction to estimate the changes their
incorporation could induce on the mean activity index and mean morphology in
each group. In Table~6, we list the 16 groups for which the activity
information is incomplete, identifying the missing galaxies and their
morphology. In column~4 we report the observed mean activity index and in
column~5 we note the change expected: a $+$ sign indicates a decrease in
activity, a $-$ sign indicates an increase in activity, and a $0$ indicates no
variation. It can be seen that including the missing galaxies would either have
no impact or reinforce our analysis. The last two columns list both the actual
mean morphological type as well as its value after taking into account the
missing galaxies. Once again, only marginal changes are see. This test suggests
that our present analysis is not compromised by the incompleteness of our
sample.

\subsection{Evolutionary sequence and spatial configuration of CGs}

In Figure~9, we compare the mean level of activity with the mean morphological
type of the galaxies in 27 HCGs. The symbols distinguish between active ($\log
|{\rm EW}_{act}|)> 0.8$) and inactive groups. Our sample of CGs clearly spans
an evolutionary sequence. We see evolved CGs, which are inactive and dominated
by elliptical galaxies, slightly less evolved groups, which look unusually
active considering the early-type morphologies of their members, and
non-evolved groups, which are active and dominated by late-type spirals and
irregular galaxies.

At this point of our analysis, it is instructive to inspect the spatial
(projected) configurations of the groups. Indeed, we find that groups that
share the same level of evolution also share the same configuration. We
identify three main configuration types. Type~A CGs are dominated by late-type
spirals and all show a loose configuration. The mean activity index for this
type is high. The prototypes are HCG~16, 23, 88 and 89 and are
shown\footnote{The finding charts shown in this article were obtained using
SkyCat, which is a tool developed by ESO's Data Management and Very Large
Telescope (VLT) Project divisions with contributions from the Canadian
Astronomical Data Center (CADC); http://archive.eso.org/skycat/} in Figure~10.
The two groups HCG~30 and 87 may also be classified as Type~A, based on their
spatial configuration, even though they are inactive. Type~B consists of
galaxies in apparent close contact, suggesting ongoing mergers. The prototypes
are HCG~40, 56, 68 and 79 (Figure~11). A similar configuration is found for
HCG~34, 37 and 67. These are all unusually active groups, considering the
morphologies of their members. The three groups HCG~22, 90 and 98 may also be
classified as Type~B, even though they are inactive. Finally, Type C is formed
by CGs that are dominated by elliptical galaxies and are inactive. The
configuration is loose and the groups are either dominated by one giant
elliptical, like in HCG~42, 62, 74 and 94 (Figure~12), or by more than one
elliptical galaxy, like in HCG~15 and 86 (Figure~13).

Although the three CGs HCG~10, 58 and 93 seem to share the same level of
evolution as groups in Type~B, their spatial configurations are more similar to
groups in Type~A, and defy, consequently any classification.

The group HCG~54, also seems somewhat special. This group, which is remarkably
active, shows the same merging configuration as CGs of Type~B, but with the
important difference that it is formed by irregular galaxies. The diagnostic
diagram, Figure~3, suggests that the galaxies in this group are low-mass,
low-metallicity HII galaxies. This is in contrast with the other CGs in our
sample, which are usually formed by massive and probably metal rich galaxies
(as judged from their positions in the diagnostic diagram). It is not certain,
therefore, that HCG 54 describes the same phenomenon as the other groups.

Finally, according to Ribeiro et al. (1998), HCG~4 was misclassified as a CG.
They showed that galaxies HCG~4e and HCG~4c weren't group members. Eliminating
them causes HCG~4 to fail the definition for a compact group according to
Hickson.

Table~7 summarizes our classification for all the groups in our sample (with
the exception of HCG~4). In this table, we also list the mean characteristics
associated with the different Types. In columns 2 and 3, a colon following the
number identifies groups with deviant or marginally deviant dynamical
characteristics (for example, HCG 34 and 37 have significantly higher velocity
dispersions than the other groups belonging to Type B.

Not considering the exceptions, the CGs in the different main configuration are
clearly separated in Figure~9. HCGs with a mean T$ > 2.2$ are all Type A
systems. Those with a mean T$< -2.2$ are all of Type C. Transition groups are
classified as Type B. In terms of evolution, we may distinguish a continuous
sequence, which goes in the sense A$\Rightarrow$B$\Rightarrow$C, where A is
less evolved than C. We emphasize that in terms of spatial configuration alone,
such an evolutionary sequence is not obvious (considering also the exceptions)
and one needs a second parameter, such as the activity index introduced in this
paper, for it to become noticeable. This may explain why the different
configuration types where not recognized before.

Using a one-way ANOVA test, we confirm the differences observed between the
characteristics of Types~A, B and C, as shown in Table~7, at a 95\% confidence
level. In Table~8, we show the results of Tukey's multiple test, that allows to
identify the source of the variance\footnote{One-way ANOVA with Tukey's post
test was performed using GraphPad Prism version 3.00 for Windows, GraphPad
Software, San Diego California USA (www.graphpad.com)}. This table reports the
P values for the difference between each pair of mean values. If the null
hypothesis is true (all the values are sampled from populations with the same
mean), then there is only a 5\% chance that any one or more comparisons will
have a P value less than 0.05. These tests confirm the different evolutionary
level of the three types defined in this work.

\subsection{The relation between galaxy activity and the dynamics of CGs}

According to the conventional scenario for the fate of CGs, also referred to
as the fast merger model (Mendes de Oliveira \& Hickson 1994; G\'omez-Flechoso
\& Dom\'{\i}nguez-Tenreiro 2001), young forming groups are expected to start
out with high velocity dispersions and large radii, which would rapidly
decrease as the galaxies merge to form one giant elliptical (this is
illustrated by the model CG2 in Figure~1 of G\'omez-Flechoso \&
Dom\'{\i}nguez-Tenreiro 2001). Assuming the groups form from spiral galaxies
that originated in the field, we would thus expect young groups to be rich in
late-type galaxies and to become gradually richer in early-type galaxies before
their final merging phase. In terms of activity, we would also expect the level
of star formation and AGN to be high at the beginning of the sequence and to
decrease gradually, maybe with a change from star formation to AGN, in the
final merger phase.

According to the conventional fast merger model for CGs, we would thus expect
to see an evolutionary sequence that would trace a very specific pattern in a
graph mapping the radius as a function of the velocity dispersion. The expected
pattern is sketched in Figure~14a. The actual observed positions are shown in
Figure~14b. One can immediately see that the expected pattern is not reflected
by the observations: CGs in Type~A, which are less evolved, should have large
radii and high velocity dispersions, whereas CGs in Type C, which are more
evolved, should have small radii and low velocity dispersions. Contrary to
expectation, Type~C CGs have radii comparable to those of Type~A and much
higher velocity dispersions (see Table~7).

In Figure~15 we compare the mean activity index and morphology with the radius
of the groups (HCG~54 was omitted). We find no relation between activity or
morphology and the radius of the group. The mean activity index and morphology
are compared with the velocity dispersion in Figure~16. We distinguish no
relation between the activity and the velocity dispersion. However, we see a
clear trend for the groups dominated by early-type galaxies to have high
velocity dispersions. In terms of evolution, we see the velocity dispersion
increasing from Type~A to C (see also Tables 7 and 8), which indicates that the
most evolved groups have the highest velocity dispersions.

The trend we observe in our sample between evolution and velocity dispersion,
seems to stem from the strong correlation between morphology and velocity
dispersion previously discovered by Hickson, Kindl \& Huchra (1988). Our
observations support their claim that the general relation between morphology
and velocity dispersion encountered in CGs is an evolutionary phenomenon linked
to the environment (for other evidence in favor of this interpretation see
Shaker et al. 1998 and Focardi \& Kelm 2002).

\section{Discussion}

The trends we observe in our sample are, at best, only partially consistent
with the conventional fast merger model for CGs. We do distinguish what could
be young forming groups (Type~A) and more evolved merging groups (Type~B).
However, the velocity dispersions and radii of these systems are contrary to
what is expected. In a sense, this negative result is somewhat discomforting,
because it suggests that the dynamical evolution of galaxies in CGs is more
complex than previously believed. However, this result doesn't have to be too surprising,
since this is what one expects if CGs are in reality part of larger scale
structures (Rose 1977; West 1989; White 1990; Diaferio, Geller \& Ramella 1994;
Governato, Tozzi \& Cavaliere 1996).

Assuming the formation and evolution processes of CGs are related
to larger scale structures, we may have found a new clue how to explain their
behavior by relating the evolutionary level of a group to its velocity
dispersion. Indeed, in dynamical terms, the velocity dispersion is generally
seen to increase with the total mass of the system (Heisler, Tremaine \&
Bahcall 1985; Perea, del Olmo \& Moles 1990). By applying this observation to
CGs, we may thus conclude that their level of evolution increases with their
mass, or more properly, with the mass of the structure with which they are
associated or where they are embedded in.

The above interpretation is supported by the three following independent
observations. First, we note in Table~7 that the mean masses of the groups, as
defined by Hickson (see Tables~4 and 5), does increase from types A to C. This
is confirmed by a one-way ANOVA test at a 95\% confidence level. In Table~8,
the results of Tukey's multiple test confirm that groups in Type~C are more
massive than those in Type~A and B. Second, we note that the number of galaxies
in structures associated with CGs generally increases with the velocity
dispersion. Looking at the number of galaxies encountered by Ribeiro et al.
(1998) in Table~4 and 5, we see that the structures associated with groups of
Type~A contain between 6 and 8 galaxies, whereas those associated with groups
in Type~C contain between 9 and 18 galaxies (the only exception being HCG~22).
The third observation is the detection of diffuse X-ray envelopes. Naturally,
one expects these envelopes to be more common or easily detected in massive
systems. Using the data in Tables 4 and 5, we see that only one non--evolved
group with low velocity dispersion (lower than 300 km s$^{-1}$) has an X-ray
envelope. This is HCG~16 (Type~A), which is unusually active (see Ribeiro et
al. 1996). Another small velocity dispersion group with an X-ray envelope is
the relatively evolved group HCG~58. But, this group may be considered
unusually active as well, since it contains the only Sy1 in our sample. All the
other X-ray envelopes are detected primarily in Type~C, HCG~15, 42, 62, 86, 90,
and Type~B groups, HCG~37, 67, 68. With the exception of HCG~90, all these
groups have velocity dispersions higher than 300 km s$^{-1}$.

Assuming that the velocity dispersion increases with the mass of the structure,
a ready interpretation of our observations is that CGs associated with massive
structures are more evolved. Theoretically, it is easy to find arguments in
favor of this interpretation. For example, if we assume that a large number of
interactions accelerates the evolution of galaxies, then we would naturally
expect galaxies in more massive and rich systems to be more evolved. However,
tidal interactions become less effective as collision velocities increase,
which then seems in contradiction with what we observe. Alternatively, if the
formation of CGs follows the formation of large scale structures then, under
certain conditions, we would expect high-density structures to collapse before
low density ones, forming primordially massive structures, and CGs associated
with such structures to be naturally older.

Our observations are consistent with the formation of structures as predicted
by the CDM model (or $\Lambda$CDM), assuming the formation of galaxies is a
biased process (Davis et al. 1985; Benson et al. 2001), that is galaxies
developing first in high-density structures. In terms of evolution, biased
galaxy formation would naturally lead to more massive systems being more
evolved (see figure one in Benson et al. 2001). If one follows the CDM model
evolution over time, galaxies are seen to form first in the high-density
structures, then structures are seen to get richer and more complex through
mergers of smaller systems. As part of the structure formation process, the
evolution of small scale entities, like CGs, must consequently be more complex
than predicted by the conventional fast merger model. This is because the
galaxies in these groups are possibly influenced not only by the other galaxy
members, but also by the neighboring structures (see Einasto et al. 2003;
Ragone et al. 2004).

As part of the structure formation process, the fate of CGs may also be
different than the one predicted by the conventional fast merger model. Since
structures continuously merge to form larger ones, the first groups to form
(Type~C), being in denser environments, would already have had time to merge with
other similarly evolved groups. The most evolved groups would thus be expected
today to form part of larger structures where they would be in dynamical
equilibrium.

As for the groups that seem to have formed more recently (Type~A), or those
that formed in the recent past (Type~B), the galaxies in these groups obviously
have not had enough time to merge. Being found in less massive environments, it
is not clear whether they will experience a complete merger of their components
or whether they will find other groups with which they can merge to form larger
structures (and reach equilibrium before completing the merger of their
members). Such a scenario may explain some of the exceptions to the trends found in our
sample, like HCG 54 (and may be HCG 31) as an example of a group in a low density
environment that is completing a merger, or HCG 10, 58 and 93, as examples of
different groups merging together to form a larger structure.

Finally, it is important to note that we can possibly distinguish between the
two hypotheses suggested to explain our observations by examining CGs at higher
redshifts. In particular, if the structure formation hypothesis is correct, one
would thus expect to find CGs at higher redshift to be associated with more
massive structures than those at lower redshift.

\section{Summary and conclusions}

We have defined a new spectroscopic index that allows us to quantify the level
of activity (AGN and star formation) in galaxies. By taking the mean of this
index in a CG and combining it with the mean morphological type of the galaxies
it proves possible to quantify the evolutionary state of these systems.

Applying our method to a sample of 27 CGs from Hickson's catalog, we have found
an evolutionary sequence, which is correlated with the projected spatial
configuration of the groups. Mapping the position of CGs with different levels
of evolution and group morphology in a diagram of radius versus velocity
dispersion, we did not observe the pattern predicted by the conventional fast
merger model. Contrary to expectation, we found that the level of evolution of
CGs increases with velocity dispersion. This trend was further shown to be
possibly connected to the masses of the CGs or to the structures in which they
are embedded. Assuming that the velocity dispersion increases with the mass,
the trend we observe would thus imply that the most evolved groups are found
within the most massive structures.

We propose two different hypotheses to explain our results. The first assumes
that the evolution of galaxies accelerates with the number of interactions in
massive structures. The other assumes that the formation of CGs follows the
formation of the large scale structure, and that massive structures develop
before less massive ones. Our observations are consistent with the formation of
structures as predicted by the CDM model (or $\Lambda$CDM), assuming the
formation of galaxies is a biased process: galaxies developing first in
high-density structures.

\acknowledgements The authors would like to thank the Time Allocation Committee
of the Observatorio Astron\'omico Nacional at San Pedro M\'artir for generous
allocations of observing time, which has made this project possible. We also
would like to thank the support personnel of the observatory, and in particular
the two night assistants Felipe Montalvo Rocha and Salvador Monroy, without
whom the observations would have been arduous, to say the least. RC is grateful
to Heinz Andernach for reading and commenting on an early version of the
manuscript. The referee, Paul Hickson, is also acknowledged for his comments
and suggestions that helped improve our analysis and discussion. This research
was supported, in part, by CONACyT grant EX-000479 and 40194.

\clearpage

\begin{deluxetable}{llcccclcc}
\tabletypesize{\scriptsize} \tablecaption{Log of the observations and
main characteristics of the targets}
\tablewidth{0pt} \tablehead{ \colhead{obs.} & \colhead{HCG}  &
\colhead{$\alpha$} & \colhead{$\delta$} & \colhead{v$_r$} & \colhead{T} & \colhead{Morph.} &
\colhead{exp.} & \colhead{ Airmass }\\
\colhead{date}   & \colhead{}  & \colhead{(2000)}
&\colhead{(2000)} &  \colhead{(km s$^{-1}$)} &  \colhead{} &
\colhead{}& \colhead{($\times 900s$)} &  \colhead{}\\
\colhead{(1)} &\colhead{(2)} &\colhead{(3)} &\colhead{(4)} &\colhead{(5)} &\colhead{(6)} &\colhead{(7)} &\colhead{(8)} &\colhead{(9)}  }
\startdata
16/10 &\phs\phn 8a  & $\phn 0$ $49$ $34.26 $ &\phs$ 23$ $34$ $40.14 $ & $16014    $&    $-5$  & E5   & 3 & 1.0 \\
16/10 &\phn    +8b  & $\phn 0$ $49$ $35.29 $ &\phs$ 23$ $35$ $28.32 $ & $15966    $&    $-2$  & S0   & 3 & 1.2 \\
16/10 &\phn    +8c  & $\phn 0$ $49$ $35.90 $ &\phs$ 23$ $35$ $02.01 $ & $17087    $&    $-2$  & S0   & 3 & 1.2 \\
16/10 &\phs\phn 8d  & $\phn 0$ $49$ $36.77 $ &\phs$ 23$ $34$ $22.80 $ & $16341    $&    $-2$  & S0   & 3 & 1.0 \\
17/10 &\phs    10a  & $\phn 1$ $26$ $21.43 $ &\phs$ 34$ $42$ $07.51 $ & $\phn5148 $&\phs$ 3$  & SBb  & 3 & 1.0 \\
17/10 &\phs    10b  & $\phn 1$ $25$ $40.33 $ &\phs$ 34$ $42$ $46.39 $ & $\phn4862 $&    $-5$  & E1   & 3 & 1.1 \\
20/10 &\phs    10c  & $\phn 1$ $26$ $18.81 $ &\phs$ 34$ $45$ $14.38 $ & $\phn4660 $&\phs$ 5$  & Sc   & 3 & 1.0 \\
20/10 &\phs    10d  & $\phn 1$ $26$ $30.84 $ &\phs$ 34$ $40$ $30.46 $ & $\phn4620 $&\phs$ 6$  & Scd  & 3 & 1.1 \\
18/10 &\phs    15a  & $\phn 2$ $07$ $53.00 $ &\phs$ 02$ $10$ $03.31 $ & $\phn6967 $&\phs$ 1$  & Sa   & 3 & 1.2 \\
19/10 &\phs    15b  & $\phn 2$ $07$ $34.09 $ &\phs$ 02$ $06$ $54.73 $ & $\phn7117 $&    $-5$  & E0   & 3 & 1.1 \\
19/10 &\phs    15c  & $\phn 2$ $07$ $39.75 $ &\phs$ 02$ $08$ $59.02 $ & $\phn7222 $&    $-5$  & E0   & 3 & 1.1 \\
19/10 &\phs    15d  & $\phn 2$ $07$ $37.49 $ &\phs$ 02$ $10$ $50.80 $ & $\phn6244 $&    $-5$  & E2   & 3 & 1.2 \\
19/10 &\phs    30a  & $\phn 4$ $36$ $18.57 $ &    $-02$ $49$ $52.67 $ & $\phn4697 $&\phs$ 5$  & SBc  & 3 & 1.2 \\
19/10 &\phs    30b  & $\phn 4$ $36$ $30.27 $ &    $-02$ $51$ $59.66 $ & $\phn4625 $&\phs$ 1$  & Sa   & 3 & 1.2 \\
20/10 &\phs    31a  & $\phn 5$ $01$ $38.74 $ &    $-04$ $15$ $34.11 $ & $\phn4042 $&\phs$ 8$  & Sdm  & 4 & 1.2 \\
17/10 &\phs    34a  & $\phn 5$ $21$ $45.93 $ &\phs$ 06$ $41$ $19.73 $ & $\phn8997 $&    $-5$  & E2   & 3 & 1.1 \\
17/10 &\phs    34c  & $\phn 5$ $21$ $48.96 $ &\phs$ 06$ $40$ $54.71 $ & $\phn9392 $&\phs$ 6$  & SBd  & 1 & 1.2 \\
08/04 &\phs    37a  & $\phn 9$ $13$ $39.44 $ &\phs$ 29$ $59$ $32.61 $ & $\phn6745 $&    $-3$  & E7   & 3 & 1.0 \\
08/04 &       +37b  & $\phn 9$ $13$ $32.60 $ &\phs$ 29$ $59$ $59.04 $ & $\phn6741 $&\phs$ 4 $ & Sbc  & 6 & 1.0 \\
08/04 &       +37c  & $\phn 9$ $13$ $37.22 $ &\phs$ 29$ $59$ $58.22 $ & $\phn7357 $&\phs$ 0 $ & S0a  & 6 & 1.0 \\
08/04 &\phs    37d  & $\phn 9$ $13$ $33.84 $ &\phs$ 30$ $00$ $51.78 $ & $\phn6207 $&\phs$ 8 $ & SBdm & 3 & 1.1 \\
10/04 &\phs    40a  & $\phn 9$ $38$ $53.52 $ &    $-04$ $50$ $56.53 $ & $\phn6628 $&    $-5 $ & E3   & 3 & 1.2 \\
10/04 &\phs    40b  & $\phn 9$ $38$ $55.04 $ &    $-04$ $51$ $57.90 $ & $\phn6842 $&    $-2 $ & S0   & 3 & 1.2 \\
10/04 &\phs    40c  & $\phn 9$ $38$ $53.20 $ &    $-04$ $51$ $33.72 $ & $\phn6890 $&\phs$ 4 $ & Sbc  & 3 & 1.3 \\
10/04 &\phs    40d  & $\phn 9$ $38$ $55.77 $ &    $-04$ $50$ $14.53 $ & $\phn6492 $&\phs$ 1 $ & SBa  & 3 & 1.3 \\
10/04 &\phs    40e  & $\phn 9$ $38$ $55.45 $ &    $-04$ $51$ $27.91 $ & $\phn6625 $&\phs$ 5 $ & Sc   & 3 & 1.2 \\
09/04 &       +54a  & $    11$ $29$ $15.17 $ &\phs$ 20$ $35$ $00.73 $ & $\phn1397 $&\phs$ 8 $ & Sdm  & 4 & 1.0 \\
09/04 &       +54b  & $    11$ $29$ $14.05 $ &\phs$ 20$ $34$ $53.24 $ & $\phn1412 $&\phs$ 10$ & Im   & 4 & 1.0 \\
09/04 &\phs    54c  & $    11$ $29$ $16.26 $ &\phs$ 20$ $35$ $11.12 $ & $\phn1420 $&\phs$ 10$ & Im   & 3 & 1.0 \\
09/04 &\phs    54d  & $    11$ $29$ $16.50 $ &\phs$ 20$ $35$ $18.82 $ & $\phn1670 $&\phs$ 10$ & Im   & 3 & 1.1 \\
08/04 &       +56a  & $    11$ $32$ $46.64 $ &\phs$ 52$ $56$ $27.03 $ & $\phn8245 $&\phs$  5$ & Sc   & 3 & 1.5 \\
08/04 &       +56b  & $    11$ $32$ $40.45 $ &\phs$ 52$ $57$ $01.59 $ & $\phn7919 $&    $ -2$ & SB0  & 5 & 1.1 \\
08/04 &\phs    56c  & $    11$ $32$ $36.69 $ &\phs$ 52$ $56$ $51.03 $ & $\phn8110 $&    $ -2$ & S0   & 3 & 1.1 \\
08/04 &\phs    56d  & $    11$ $32$ $35.29 $ &\phs$ 52$ $56$ $49.84 $ & $\phn8346 $&    $ -2$ & S0   & 3 & 1.2 \\
08/04 &\phs    56e  & $    11$ $32$ $32.73 $ &\phs$ 52$ $56$ $20.96 $ & $\phn7924 $&    $ -2$ & S0   & 3 & 1.3 \\
10/04 &\phs    58a  & $    11$ $42$ $11.06 $ &\phs$ 10$ $16$ $39.62 $ & $\phn6138 $&\phs$  3$ & Sb   & 3 & 1.1 \\
10/04 &\phs    58b  & $    11$ $42$ $23.56 $ &\phs$ 10$ $15$ $51.05 $ & $\phn6503 $&\phs$  2$ & SBab & 3 & 1.1 \\
10/04 &\phs    58c  & $    11$ $41$ $53.15 $ &\phs$ 10$ $18$ $14.63 $ & $\phn6103 $&\phs$  0$ & SB0a & 3 & 1.2 \\
10/04 &\phs    58d  & $    11$ $42$ $05.89 $ &\phs$ 10$ $21$ $03.06 $ & $\phn6270 $&    $ -5$ & E1   & 3 & 1.2 \\
10/04 &\phs    58e  & $    11$ $42$ $04.82 $ &\phs$ 10$ $23$ $01.86 $ & $\phn6052 $&\phs$  4$ & Sbc  & 3 & 1.4 \\
09/04 &\phs    68a  & $    13$ $53$ $26.53 $ &\phs$ 40$ $16$ $58.35 $ & $\phn2162 $&    $ -2$ & S0   & 3 & 1.0 \\
09/04 &\phs    68b  & $    13$ $53$ $26.62 $ &\phs$ 40$ $18$ $08.65 $ & $\phn2635 $&    $ -5$ & E2   & 3 & 1.0 \\
09/04 &\phs    68c  & $    13$ $53$ $21.69 $ &\phs$ 40$ $21$ $47.68 $ & $\phn2313 $&\phs$  4$ & SBbc & 4 & 1.1 \\
08/04 &\phs    74a  & $    15$ $19$ $24.91 $ &\phs$ 20$ $53$ $44.17 $ & $12255    $&    $ -5$ & E1   & 3 & 1.1 \\
09/04 &\phs    74b  & $    15$ $19$ $24.38 $ &\phs$ 20$ $53$ $23.44 $ & $12110    $&    $ -5$ & E3   & 3 & 1.1 \\
09/04 &\phs    74c  & $    15$ $19$ $26.18 $ &\phs$ 20$ $53$ $56.44 $ & $12266    $&    $ -2$ & S0   & 3 & 1.1 \\
09/04 &\phs    74d  & $    15$ $19$ $32.02 $ &\phs$ 20$ $52$ $57.47 $ & $11681    $&    $ -5$ & E2   & 3 & 1.2 \\
07/04 &\phs    79a  & $    15$ $59$ $11.41 $ &\phs$ 20$ $45$ $14.86 $ & $\phn4292 $&    $ -5$ & E0   & 3 & 1.0 \\
07/04 &\phs    79b  & $    15$ $59$ $12.61 $ &\phs$ 20$ $45$ $47.14 $ & $\phn4446 $&    $ -2$ & S0   & 3 & 1.0 \\
07/04 &\phs    79c  & $    15$ $59$ $10.95 $ &\phs$ 20$ $45$ $41.43 $ & $\phn4146 $&    $ -2$ & S0   & 2 & 1.0 \\
10/04 &\phs    79d  & $    15$ $59$ $12.01 $ &\phs$ 20$ $44$ $47.20 $ & $\phn4503 $&\phs$ 8 $ & Sdm  & 3 & 1.1 \\
17/10 &\phs    88b  & $    20$ $52$ $29.69 $ &   $ -05$ $44$ $47.61 $ & $\phn6010 $&\phs$ 3 $ & SBb  & 4 & 1.2 \\
17/10 &\phs    88c  & $    20$ $52$ $25.94 $ &   $ -05$ $46$ $20.12 $ & $\phn6083 $&\phs$ 5 $ & Sc   & 2 & 1.3 \\
16/10 &\phs    89a  & $    21$ $20$ $01.03 $ &   $ -03$ $55$ $20.26 $ & $\phn8850 $&\phs$ 5 $ & Sc   & 3 & 1.2 \\
16/10 &\phs    89b  & $    21$ $20$ $19.21 $ &   $ -03$ $53$ $46.81 $ & $\phn8985 $&\phs$  5$ & SBc  & 3 & 1.5 \\
17/10 &\phs    93a  & $    23$ $15$ $15.98 $ &\phs$ 18$ $57$ $41.36 $ & $\phn5140 $&    $ -5$ & E1   & 3 & 1.0 \\
17/10 &\phs    93b  & $    23$ $15$ $17.17 $ &\phs$ 19$ $02$ $29.77 $ & $\phn4672 $&\phs$  6$ & SBd  & 3 & 1.0 \\
17/10 &\phs    93c  & $    23$ $15$ $03.59 $ &\phs$ 18$ $58$ $23.28 $ & $\phn5132 $&\phs$  1$ & SBa  & 3 & 1.1 \\
20/10 &\phs    93d  & $    23$ $15$ $33.10 $ &\phs$ 19$ $02$ $52.50 $ & $\phn5173 $&    $ -2$ & SB0  & 3 & 1.0 \\
18/10 &       +94a  & $    23$ $17$ $13.44 $ &\phs$ 18$ $42$ $28.22 $ & $12040    $&    $ -5$ & E1   & 3 & 1.0 \\
18/10 &       +94b  & $    23$ $17$ $11.94 $ &\phs$ 18$ $42$ $03.30 $ & $11974    $&    $ -5$ & E3   & 3 & 1.0 \\
18/10 &       +94c  & $    23$ $17$ $20.26 $ &\phs$ 18$ $44$ $03.62 $ & $12120    $&    $ -2$ & S0   & 3 & 1.0 \\
20/10 &\phs    94d  & $    23$ $17$ $15.19 $ &\phs$ 18$ $42$ $42.55 $ & $13009    $&    $ -2$ & S0   & 3 & 1.1 \\
18/10 &       +98a  & $    23$ $54$ $10.08 $ &\phs$ 00$ $22$ $58.07 $ & $\phn7855 $&    $ -2$ & SB0  & 5 & 1.2 \\
18/10 &       +98b  & $    23$ $54$ $12.21 $ &\phs$ 00$ $22$ $35.88 $ & $\phn7959 $&    $ -2$ & S0   & 5 & 1.2 \\
\enddata
\end{deluxetable}

\clearpage
\begin{deluxetable}{cccccclr}
\tabletypesize{\scriptsize} \tablecaption{Observed line ratios, fluxes and Equivalent Widths}
\tablewidth{0pt} \tablehead{
\colhead{HCG}  & \colhead{Ap.} & \colhead{$\log$([NII]/H$\alpha$)}& \colhead{$\log$([OIII]/H$\beta$)} &
\colhead{$\log($L$_{{\rm H}\alpha})$} & \colhead{EW(H$\alpha+{\rm NII}$)} & \colhead{Act. type} & \colhead{EW$_{\rm act}$} \\
\colhead{} & \colhead{(kpc)} & \colhead{}& \colhead{} & \colhead{(erg s$^{-1}$)}& \colhead{(\AA)} & \colhead{} & \colhead{(\AA)}\\
\colhead{(1)} &\colhead{(2)} &\colhead{(3)} &\colhead{(4)} &\colhead{(5)} &\colhead{(6)} &\colhead{(7)} &\colhead{(8)}
}
\startdata
\phn 8a & 12 &\nodata             &\nodata            &\nodata &\nodata & No em.   &\nodata    \\
\phn 8b & 11 &\nodata             &\nodata            &\nodata &\nodata & No em.   &\nodata    \\
\phn 8c & 10 &\nodata             &\nodata            &\nodata &\nodata & No em.   &\nodata    \\
\phn 8d & 11 &\nodata             &\nodata            &\nodata &\nodata & No em.   &\nodata    \\
    10a &  4 &\phs$0.39\pm 0.10 $ &\phs$0.76\pm 0.17$ & 39.4:  & $-5 $  & dSy2   &   $-2.8$  \\
    10b &  3 &\nodata             &\nodata            &\nodata &\nodata & No em.   &\phs$2.6$  \\
    10c &  3 & $-0.33\pm 0.02$    &\nodata            & 39.6   & $-15$  & SFG?   & $ -12.5$  \\
    10d &  3 & $-0.29\pm 0.03$    &\nodata            & 39.0   & $-10$  & SFG?   & $  -9.4$  \\
    15a &  4 & $-0.01\pm 0.08$    & $ -0.19\pm 0.15$  & 40.0   & $-3 $  & dLINER & $   0.3$  \\
    15b &  4 &\nodata             &\nodata            &\nodata &\nodata & No em.   & $   4.4$  \\
    15c &  4 &\nodata             &\nodata            &\nodata &\nodata & No em.   & $   5.7$  \\
    15d &  4 & $-0.02\pm 0.04$    &\phs$0.11\pm 0.06$ & 39.8   & $-9 $  & LINER  & $  -8.6$  \\
    30a &  3 &\nodata             &\nodata            &\nodata &\nodata & No em.   & $   3.8$  \\
    30b &  2 &\nodata             & $-0.34\pm 0.16$   &\nodata &\nodata & LLAGN?  & $   2.4$  \\
 $+$31a &  3 & $-0.95\pm 0.02$    &\phs$0.44\pm 0.02$ & 40.2   &$-693$  & HIIgal & $-592.1$  \\
    34a &  5 &\phs$0.02\pm 0.08$  &\nodata            & 39.8:  & $-3 $  & LLAGN?  & $   1.7$  \\
 $+$34c &  6 & $-0.52\pm 0.05$    & $ -0.02\pm 0.06$  & 39.6   & $-97$  & SFG    & $-100.4$  \\
    37a &  7 &\phs$0.23\pm 0.06$  &\phs$0.22\pm 0.22$ & 40.0:  & $-5 $  & dLINER & $  -2.9$  \\
    37b & 13 & $-0.19\pm 0.07$    &\phs$0.10\pm 0.16$ & 39.7   & $-5 $  & LINER  & $  -0.6$  \\
    37c & 10 & $-0.28\pm 0.10$    &\nodata            & 39.2:  & $-2 $  & LLAGN?  & $   0.3$  \\
    37d &  6 & $-0.57\pm 0.01$    & $ -0.46\pm 0.08$  & 40.1   & $-44$  & SFG    & $ -42.2$  \\
    40a &  5 &\nodata             &\nodata            &\nodata &\nodata & No em.   & $   1.0$  \\
    40b &  6 &\nodata             &\nodata            &\nodata &\nodata & No em.   & $   3.1$  \\
    40c &  8 & $-0.23\pm 0.01$    & $ -0.30\pm 0.09$  & 40.2   & $-14$  & SFG    & $ -12.0$  \\
    40d &  5 & $-0.28\pm 0.01$    & $ -0.46\pm 0.08$  & 40.8   & $-38$  & SFG    & $ -40.1$  \\
    40e &  6 & $-0.31\pm 0.01$    &\nodata            & 39.4   & $-6 $  & SFG?   & $  -3.2$  \\
    54a &  1 & $-0.90\pm 0.03$    &\phs$0.51\pm 0.06$ & 38.7   & $-18$  & HIIgal & $ -14.4$  \\
 $+$54b &  1 & $-1.36\pm 0.03$    &\phs$0.68\pm 0.03$ & 38.5   &$-471$  & HIIgal & $-472.4$  \\
 $+$54c &  1 & $-1.12\pm 0.02$    &\phs$0.37\pm 0.03$ & 37.8   & $-84$  & HIIgal & $ -78.5$  \\
 $+$54d &  1 & $-1.23\pm 0.05$    &\phs$0.57\pm 0.05$ & 37.8   &$-102$  & HIIgal & $-102.7$  \\
    56a & 25 & $-0.42\pm 0.01$    &\phs$0.01\pm 0.04$ & 40.4   & $-20$  & SFG    & $ -16.8$  \\
 $+$56b &  5 & $-0.04\pm 0.01$    &\phs$1.45\pm 0.02$ & 39.9   & $-52$  & Sy2    & $ -52.1$  \\
    56c &  6 &\nodata             &\nodata            &\nodata &\nodata & No em.   & $   1.4$  \\
    56d & 11 & $-0.28\pm 0.02$    & $ -0.10\pm 0.02$  & 40.3   & $-25$  & SFG    & $ -23.9$  \\
    56e &  7 & $-0.48\pm 0.02$    &\nodata            & 39.9   & $-14$  & SFG?   & $  -9.4$  \\
 $+$58a &  4 &\nodata             &\nodata            & 39.6   & $-66$  & Sy1.8  & $ -39.8$  \\
    58b &  5 &\phs$0.07\pm 0.18$  &\nodata            & 39.8:  & $-3 $  & LLAGN?  & $   0.2$  \\
    58c &  5 &\phs$0.10\pm 0.11$  &\phs$0.15\pm 0.16$ & 39.7:  & $-6 $  & dLINER & $  -4.6$  \\
    58d &  5 &\nodata             &\nodata            &\nodata &\nodata & No em.   & $   2.2$  \\
    58e &  5 & $-0.54\pm 0.02$    & $ -0.49\pm 0.15$  & 40.0   & $-28$  & SFG    & $ -22.4$  \\
    68a &  3 &\nodata             &\nodata            &\nodata &\nodata & No em.   & $   0.9$  \\
    68b &  2 & $-0.19\pm 0.2 $    &\phs$0.14\pm 0.17$ & 39.4:  & $-4 $  & dLINER & $   1.1$  \\
    68c &  2 & $-0.06\pm 0.02$    & $ -0.02\pm 0.05$  & 39.8   & $-31$  & LINER  & $ -29.6$  \\
    74a & 14 &\nodata             &\nodata            &\nodata &\nodata & No em.   & $   1.3$  \\
    74b & 11 &\nodata             &\nodata            &\nodata &\nodata & No em.   & $   2.5$  \\
    74c &  7 &\nodata             &\nodata            &\nodata &\nodata & No em.   & $   2.8$  \\
    74d &  9 &\nodata             &\nodata            &\nodata &\nodata & No em.   & $   2.2$  \\
    79a &  4 & $-0.28\pm 0.05$    &\phs$0.28\pm 0.11$ & 39.6   & $-7 $  & LINER  & $  -4.6$  \\
    79b &  5 & $-0.31\pm 0.05$    &\nodata            & 40.0   & $-8 $  & SFG?   & $  -6.0$  \\
    79c &  3 &\nodata             &\nodata            &\nodata &\nodata & No em.   & $   5.0$  \\
 $+$79d &  5 &$-1.05\pm 0.05$     &\phs$0.39\pm 0.04$ & 38.3   & $-47$  & HIIgal & $ -34.6$  \\
    88b &  3 &\phs$0.28\pm 0.09$  &\phs$0.51\pm 0.20$ & 39.4:  & $-7 $  & dSy2   & $  -6.1$  \\
    88c &  6 & $-0.50 \pm 0.01$   & $ -0.41\pm 0.04$  & 40.0   & $-30$  & SFG    & $ -29.9$  \\
    89a &  5 & $-0.40 \pm 0.02$   & $ -0.37\pm 0.07$  & 39.8   & $-10$  & SFG    & $  -8.7$  \\
    89b &  6 & $-0.56\pm 0.01$    & $ -0.10 \pm 0.04$ & 40.3   & $-39$  & SFG    & $ -31.3$  \\
    93a &  3 &\phs$0.13\pm 0.11$  &\phs$0.24\pm 0.14$ & 39.7   & $-7 $  & dLINER & $  -5.4$  \\
    93b &  2 & $-0.39\pm 0.02$    &\nodata            & 39.6   & $-22$  & SFG?   & $ -16.4$  \\
    93c &  3 & $-0.09\pm 0.10$    & $ -0.12\pm 0.17$  & 39.3   & $-4 $  & dLINER & $  -1.8$  \\
    93d &  3 &\nodata             &\nodata            &\nodata &\nodata & No em.   & $   3.8$  \\
    94a &  6 &\nodata             &\nodata            &\nodata &\nodata & No em.   & $   4.5$  \\
    94b &  6 &\nodata             &\nodata            &\nodata &\nodata & No em.   & $   4.3$  \\
    94c &  7 &\nodata             &\nodata            &\nodata &\nodata & No em.   & $   6.1$  \\
    94d &  6 &\nodata             &\nodata            &\nodata &\nodata & No em.   & $   6.0$  \\
    98a &  5 &\nodata             &\nodata            &\nodata &\nodata & No em.   & $   8.6$  \\
    98b &  4 &\nodata             &\nodata            &\nodata &\nodata & No em.   & $   3.0$  \\
\enddata
\end{deluxetable}

\clearpage
\begin{deluxetable}{lcccc}
\tabletypesize{\scriptsize} \tablecaption{Activity classification
in CGs} \tablewidth{0pt} \tablehead{ \colhead{} & \colhead{No em.}
& \colhead{LLAGN} & \colhead{AGN}& \colhead{SFG} } \startdata
This work (65gal/19gr)   & 38\% &  19\% & 9\% & 34\% \\
HCG (62gal/17gr)   & 40\% &  21\% &18\% & 21\% \\
SCG (193gal/49gr)  &   27\% & 22\% &19\% & 32\%\\
\enddata
\end{deluxetable}

\clearpage
\begin{deluxetable}{lllrr}
\tabletypesize{\scriptsize} \tablecaption{Activity indices and
characteristics of the additional sample}
\tablewidth{0pt} \tablehead{ \colhead{HCG} & \colhead{Act. Type\tablenotemark{a}}
& \colhead{Morph.\tablenotemark{b} }& \colhead{T} & \colhead{EW$_{\rm act}$} }
\startdata
4a & SFG   & Sc   &   5  & -127.0 \\
4b & SFG   & Sc   &   5  & -431.8 \\
4d & SFG   & E4   &  -5  &  -38.4 \\
16a& LINER & SBab &   2  &   -8.2 \\
16b& Sy2   & Sab  &   2  &  -12.5 \\
16c& SFG   & Im   &  10  & -180.2 \\
16d& SFG   & Im   &  10  &  -63.1 \\
22a& LLAGN & E2   &  -5  &    0.8 \\
22b& No Em.& Sa   &   1  &    5.1 \\
22c& SFG   & SBcd &   6  &   -0.4 \\
23a& LLAGN & Sab  &   2  &   -0.1 \\
23c& LLAGN & S0   &  -2  &    0.4 \\
23d& SFG   & Sd   &   7  & -219.8 \\
42a& LLAGN & E3   &  -5  &   -1.0 \\
42b& No Em.& SB0  &  -2  &    2.8 \\
42c& No Em.& E2   &  -5  &    3.1 \\
62a& LLAGN & E3   &  -5  &   -2.5 \\
62b& No Em.& S0   &  -2  &    3.5 \\
62c& No Em.& S0   &  -2  &    3.1 \\
67a& No Em.& E1   &  -5  &    3.6 \\
67b& SFG   & Sc   &   5  &  -19.0 \\
86a& LLAGN & E2   &  -5  &    1.2 \\
86b& LLAGN & E2   &  -5  &   -1.2 \\
86c& LINER & SB0  &  -2  &   -2.7 \\
86d& No Em.& S0   &  -2  &    2.7 \\
87a& LLAGN & Sbc  &   4  &    0.1 \\
87b& LLAGN & S0   &  -2  &    1.6 \\
90a& Sy2   & Sa   &   1  &   -3.4 \\
90b& LINER & E0   &  -2  &  -10.1 \\
90c& No Em.& E0   &  -2  &    2.2 \\
90d& LINER & Im   &  10  &   -8.4 \\
\enddata
\tablenotetext{a}{Coziol et al. 1998a}
\tablenotetext{b}{``Centre de Donn\'ees astronomiques de
Strasbourg'' (CDS), CatalogVII/213}
\end{deluxetable}

\clearpage
\begin{deluxetable}{clccccccccc}
\tabletypesize{\scriptsize} \tablecaption{Mean activity index and
morphology of CGs} \tablewidth{0pt} \tablehead{
\colhead{HCG} & \colhead{z\tablenotemark{a}} & \colhead{diff.
X\tablenotemark{b}} & \colhead{N$_S$}& \colhead{N$_{H}$\tablenotemark{a}}
& \colhead{N$_{R}$\tablenotemark{c}}&
\colhead{vel. disp.\tablenotemark{a}}&
\colhead{radius\tablenotemark{a}}&\colhead{log(Mass)\tablenotemark{a}}& \colhead{$\langle$T$\rangle$}
& \colhead{$\langle {\rm EW}_{act}\rangle$}\\
\colhead{}& \colhead{}& \colhead{}& \colhead{}& \colhead{}&
\colhead{}& \colhead{(km s$^{-1}$)}& \colhead{(kpc)}& \colhead{(g)}& \colhead{}&
\colhead{} \\
\colhead{(1)} &\colhead{(2)} &\colhead{(3)} &\colhead{(4)} &\colhead{(5)} &\colhead{(6)} &\colhead{(7)} &\colhead{(8)}
&\colhead{(9)}&\colhead{(10)}&\colhead{(11)}
}
\startdata
04 & 0.0280 & ?      & 3 &\nodata &\nodata   &    $575$ & $58$ &46.45 & \phs$ 1.7 $&$                -199$ \\
10 & 0.0161 &\nodata& 4 &\nodata&\nodata& 363 &  93   &46.09   & \phs 2.2  & \phn\phn$       -4$ \\
15 & 0.0228 &yes    & 4 & 6     &\nodata& 724 &  78   &46.86   &    $-3.5$ & \phs\phn\phn$    0$ \\
16 & 0.0132 & yes    & 4 &\nodata &\phn$  7$ &    $204$ & $45$ &45.41 & \phs$ 6.0 $&\phn$             -66$ \\
22 & 0.0090 &\nodata & 3 &\nodata &\phn$  4$ &\phn  18  & $27$ &43.70 & \phs$ 0.7 $&\phs\phn\phn$       2$ \\
23 & 0.0161 &\nodata & 3 & 4      &\phn$  8$ &    $275$ & $66$ &45.95 & \phs$ 2.3 $&\phn$             -73$ \\
30 & 0.0154 &\nodata& 2 & 4     &\nodata& 110 &  51   &44.96   & \phs 3.0  & \phs\phn\phn$    3$ \\
34 & 0.0307 &\nodata& 2 & 4     &\nodata& 550 & 15    &45.90   & \phs 0.5  & \phn$          -49$ \\
37 & 0.0223 &yes    & 4 & 5     &\nodata& 692 & 29    &46.25   & \phs 2.2  & \phn$          -11$ \\
40 & 0.0223 &\nodata& 5 & 5     & 7     & 251 & 15    &45.26   & \phs 0.6  & \phn$          -10$ \\
42 & 0.0133 & yes    & 3 & 4      &$ 11$     &    $363$ & $45$ &45.83 & $    -4.0    $&\phs\phn\phn$    2$ \\
54 & 0.0049 &\nodata& 4 &\nodata&\nodata& 182 &\phn 2 &43.88   & \phs 9.5  & $             -167$ \\
56 & 0.0270 &\nodata& 5 &\nodata&\nodata& 282 & 21    &45.34   &    $-0.6$ & \phn$          -20$ \\
58 & 0.0207 &yes    & 5 &\nodata&\nodata& 275 & 89    &45.98   & \phs 0.8  & \phn$          -13$ \\
62 & 0.0137 & yes    & 3 & 4      &\nodata   &    $490$ & $27$ &45.87 & $    -3.0    $&\phs\phn\phn$    1$ \\
67 & 0.0245 & yes    & 2 & 4      & 14       &    $363$ & $49$ &46.03 & \phs$ 0.0 $&\phs\phn$          -8$ \\
68 & 0.0080 &yes    & 3 & 5     &\nodata& 263 & 33    &45.47   &    $-1.0$ & \phn\phn$       -9$ \\
74 & 0.0399 &\nodata& 4 & 5     &\nodata& 537 & 39    &46.32   &    $-4.2$ & \phs\phn\phn$    2$ \\
79 & 0.0145 &\nodata& 4 &\nodata&\nodata& 229 &\phn 7 &44.83   &    $-0.2$ & \phn$          -10$ \\
86 & 0.0199 & yes    & 4 &\nodata &\nodata   &    $457$ & $47$ &46.20 & $    -3.5    $&\phs\phn\phn$    0$ \\
87 & 0.0296 &\nodata & 3 &\nodata &\phn$  6$ &    $145$ & $31$ &44.91 & \phs$ 3.0 $&\phs\phn\phn$       1$ \\
88 & 0.0201 &\nodata& 2 & 4 & 6     &\phn  27 & 68    &\nodata & \phs 4.0  & \phn$          -18$ \\
89 & 0.0297 &\nodata& 2 & 4&\nodata&\phn   52 & 59    &44.86   & \phs 5.0  & \phn$          -20$ \\
90 & 0.0088 & yes    & 4 &\nodata &\phn$  9$ &    $166$ & $30$ &45.00 & \phs$ 1.8 $&\phs\phn$          -5$ \\
93 & 0.0168 &\nodata& 4 &\nodata&\nodata& 355 & 71    &46.10   & \phs 0.0  & \phn\phn$       -5$ \\
94 & 0.0417 &\nodata& 5 & 7     &\nodata& 832 & 58    &46.72   &    $-3.5$ & \phs\phn\phn$    5$ \\
98 & 0.0266 &\nodata& 2 & 3     &\nodata& 204 & 28    &45.35   &    $-2.0$ & \phs\phn\phn$    6$ \\
\enddata
\tablenotetext{a}{``Centre de Donn\'ees astronomiques de
Strasbourg'' (CDS), CatalogVII/213} \tablenotetext{b}{Ponman et
al. 1996}\tablenotetext{c}{Ribeiro et al. 1998}
\end{deluxetable}

\clearpage
\begin{deluxetable}{cllcccc}
\tabletypesize{\scriptsize} \tablecaption{Effect expected of
incompleteness on our analysis} \tablewidth{0pt} \tablehead{
\colhead{HCG} & \colhead{Missing gal.} & \colhead{Morph.} &
\colhead{$\langle {\rm EW}_{\rm act}\rangle$}& \colhead{changes}&
\colhead{$\langle$T$\rangle$}& \colhead{$\langle$T$\rangle_C$}}
\startdata
04 & c       & E2        &$ -199$             &$ + $& \phs$  1.7$ &\phs$  0.0$ \\
15 & e,f     & Sa, Sbc   &\phs\phn\phn$0$     &$ 0 $&     $ -2.4$ &    $ -0.1$ \\
23 & b,      & Sc        &\phn$  -73$         &$ - $& \phs$  2.3$ &\phs$  3.0$ \\
30 & c, d    & SBc, S0   &\phs\phn\phn$3$     &$ 0 $& \phs$  3.0$ &\phs$  2.0$ \\
%31 & b, c    & Sm, Im    &$ -592$             &$ - $& \phs$  8.0$ &\phs$  8.0$ \\
34 & b, d    & Sd, S0    &\phn$  -49$         &$ 0 $& \phs$  0.5$ &\phs$  1.5$ \\
37 & e       & E0        &\phn$  -11$         &$ + $& \phs$  2.2$ &\phs$  1.4$ \\
40 & c       & Sbc       &\phn$  -10$         &$ - $&     $ -0.2$ &\phs$  0.6$ \\
42 & d       & E         &\phs\phn\phn$    2$ &$ + $&     $ -4.0$ &    $ -4.2$ \\
62 & d       & E         &\phs\phn\phn$    1$ &$ + $&     $ -3.0$ &    $ -3.5$ \\
67 & c, d    & Scd, S0   &\phs\phn$   -8$     &$ 0 $& \phs$  0.0$ &\phs$  1.0$ \\
68 & de, e   & E, S0     &\phn\phn$   -9$     &$ + $&     $ -1.0$ &    $ -2.0$ \\
74 & e       & S0        &\phs\phn\phn$    2$ &$ + $&     $ -4.2$ &    $ -3.8$ \\
88 & a, d    & Sbc, Sc   &\phn$  -18$         &$ - $& \phs$  4.0$ &\phs$  4.2$ \\
89 & c, d    & Scd, Sm   &\phn$  -20$         &$ - $& \phs$  5.0$ &\phs$  6.2$ \\
94 & e, f, g & Sd, S0, S0&\phs\phn\phn$    5$ &$ + $&     $ -3.5$ &    $ -1.6$ \\
98 & c       & E         &\phs\phn\phn$    6$ &$ + $&     $ -2.0$ &    $ -0.8$ \\

\enddata
\end{deluxetable}

\clearpage
\begin{deluxetable}{cccccccc}
\tabletypesize{\scriptsize} \tablecaption{Configuration Types and mean properties}
\tablewidth{0pt} \tablecolumns{6}
\tablehead{
\colhead{Type} & \multicolumn{2}{c}{HCG group \# }    &
\colhead{$\langle{\rm Vel. disp.}\rangle$}    & \colhead{$\langle{\rm Radius}\rangle$} &
\colhead{$\langle{\rm M}\rangle$} &
\colhead{$\langle{\rm T}\rangle$} & \colhead{$\langle{\rm EW}\rangle$}\\
\colhead{}     & \colhead{active} &\colhead{inactive} &
\colhead{(km s$^{-1}$)} & \colhead{(kpc)}  &\colhead{(g)} &\colhead{}  &\colhead{(\AA)}\\
\colhead{(1)} &\colhead{(2)} &\colhead{(3)} &\colhead{(4)} &\colhead{(5)} &\colhead{(6)} &\colhead{(7)} &\colhead{(8)}
}
\startdata
A & 16, 23, 88, 89 & 30, 87                   & $135\pm38$& $   53\pm6$&$45.2\pm.2$   & \phs$ 3.9\pm0.6$ & $ -29\pm    13$ \\
B & 37:, 34:, 40, 56, 67, 68, 79 & 22, 90, 98 & $302\pm61$& $   25\pm4$&$45.3\pm.2$   & \phs$ 0.2\pm0.4$ & $ -11\pm\phn 5$ \\
C & \nodata        &15, 42, 62:, 86, 74, 94   & $567\pm72$& $   49\pm7$&$46.2\pm.2$   & $    -3.6\pm0.2$ & \phs$\phn 2\pm\phn  1$ \\
exceptions& 10, 54, 4          & 58, 93       & \nodata   & \nodata    &\nodata       &\nodata           & \nodata \\
\enddata
\end{deluxetable}

\clearpage
\begin{deluxetable}{cccccccc}
\tabletypesize{\scriptsize} \tablecaption{Results for Tukey's
post tests} \tablewidth{0pt} \tablehead{
\colhead{Types} & \colhead{$\langle{\rm Vel. disp.}\rangle$}    &
\colhead{$\langle{\rm Radius}\rangle$} & \colhead{$\langle{\rm M}\rangle$} & \colhead{$\langle{\rm
T}\rangle$} & \colhead{$\langle{\rm EW}\rangle$}
} \startdata
A-B & $>0.05$ &$<0.01$&$>0.05$ &$<0.001$&$>0.05$\\
A-C & $<0.001$&$>0.05$&$<0.05$ &$<0.001$&$<0.05$\\
B-C & $<0.05$ &$<0.05$&$<0.05$ &$<0.001$&$>0.05$\\
\enddata
\end{deluxetable}

\clearpage

\begin{figure}
%\epsscale{0.80}
\plottwo{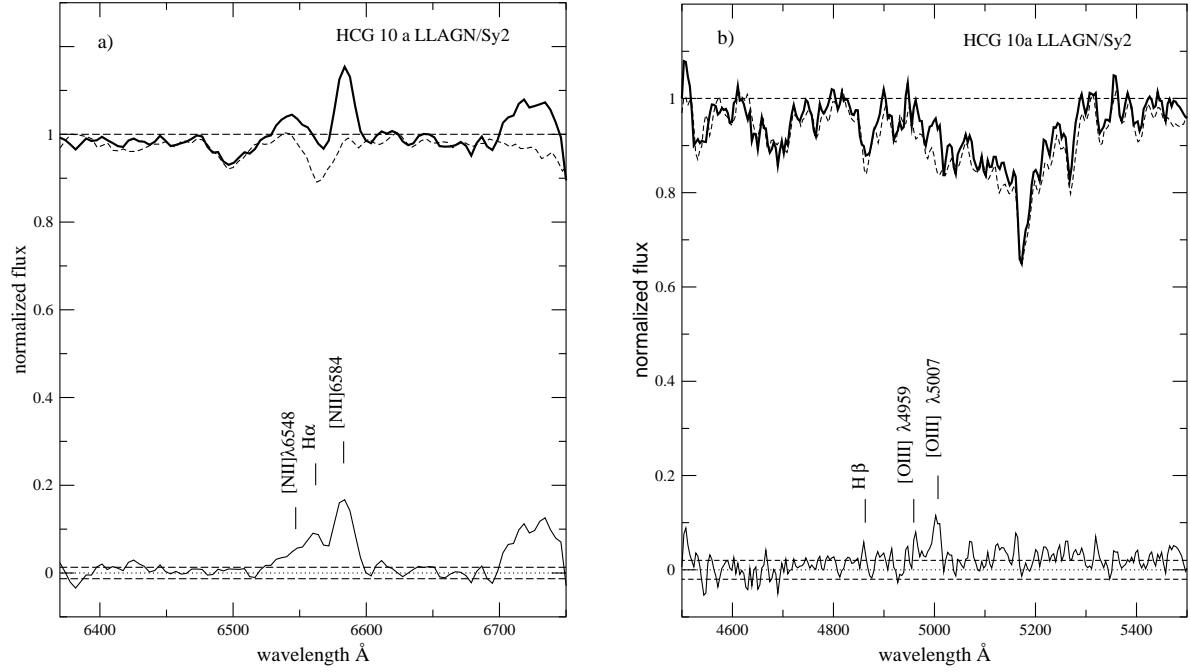}{rcoziol.fig1b.eps} \caption{Examples of template
subtraction in (a) the H$\alpha$ and (b) the H$\beta$ regions. The galaxy is
the LLAGN candidate HCG10a. The top of the figures shows the normalized spectra
before subtraction (solid line: object; dashed line: the template). The bottom
of the figures shows the residual spectra after subtraction of the template,
with the zero level (dotted) and $\pm 1 \sigma$ rms (dashed) indicated. }
\end{figure}

\clearpage

\begin{figure}
\epsscale{0.80} \plotone{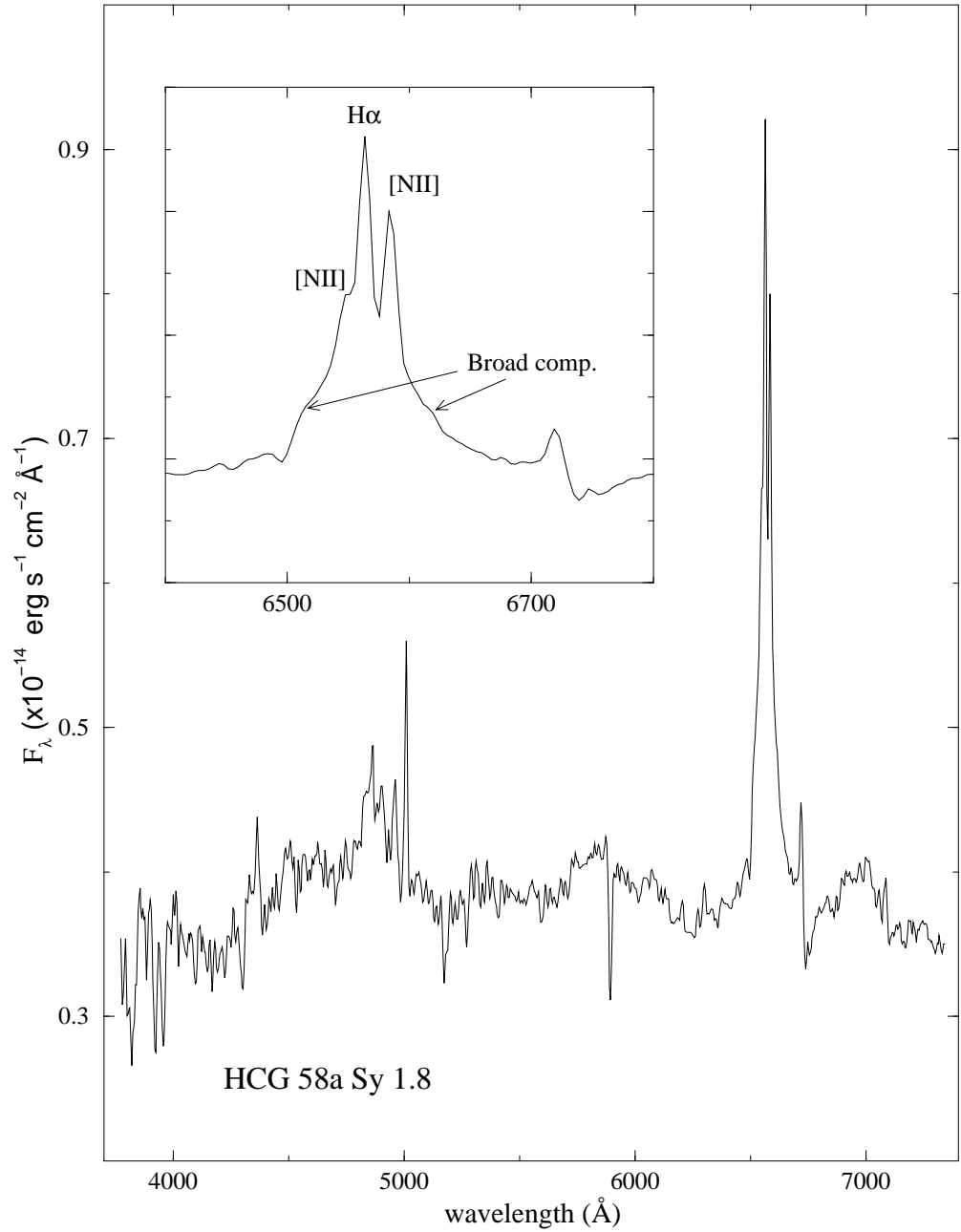} \caption{Optical spectrum of HCG
58a, presented in its restframe. This galaxy looks like a Seyfert 2, but the
presence of large wings around H$\beta$ (already evident in the spectrum) and
H$\alpha$ (evident in the enlarged section) make it a Seyfert~1.8. }
\end{figure}

\clearpage

\begin{figure}
\epsscale{0.70} \plotone{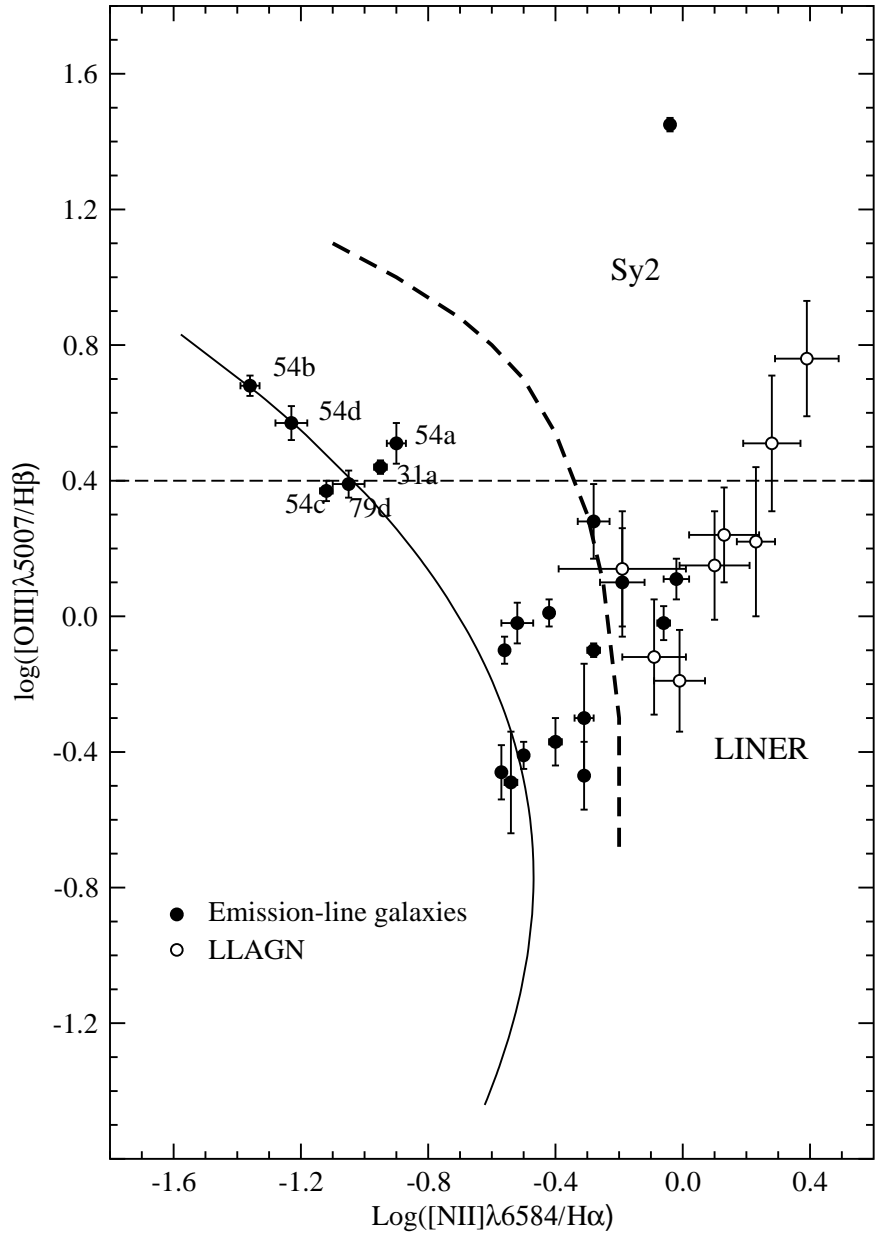} \caption{Standard diagnostic diagram
for the emission-line galaxies in our sample. The dashed curve separates
thermal (left) from non-thermal (right) galaxies, while the continuous curve is
the locus traced by normal HII regions. The horizontal dashed line separates
low excitation, high metallicity star forming galaxies (bottom) from high
excitation, low metallicity star forming galaxies (top).}
\end{figure}

\clearpage

\begin{figure}
\epsscale{0.9} \figurenum{4} \plottwo{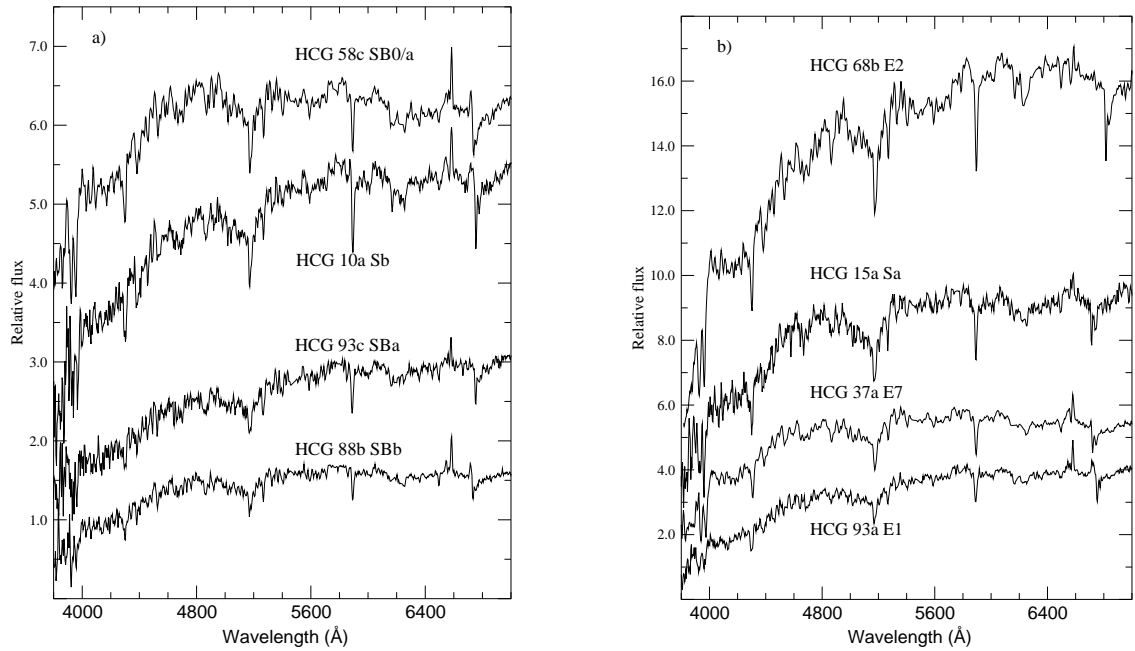}{rcoziol.fig4b.eps}
\caption{Spectra of the confirmed LLAGNs in our sample. All spectra are in the
restframe, and slightly shifted in flux to facilitate comparison. Note the
weakness of the H$\alpha$ line in emission compared to the two [NII] lines.}
\end{figure}

\clearpage

\begin{figure}
\epsscale{0.9}\figurenum{5} \plottwo{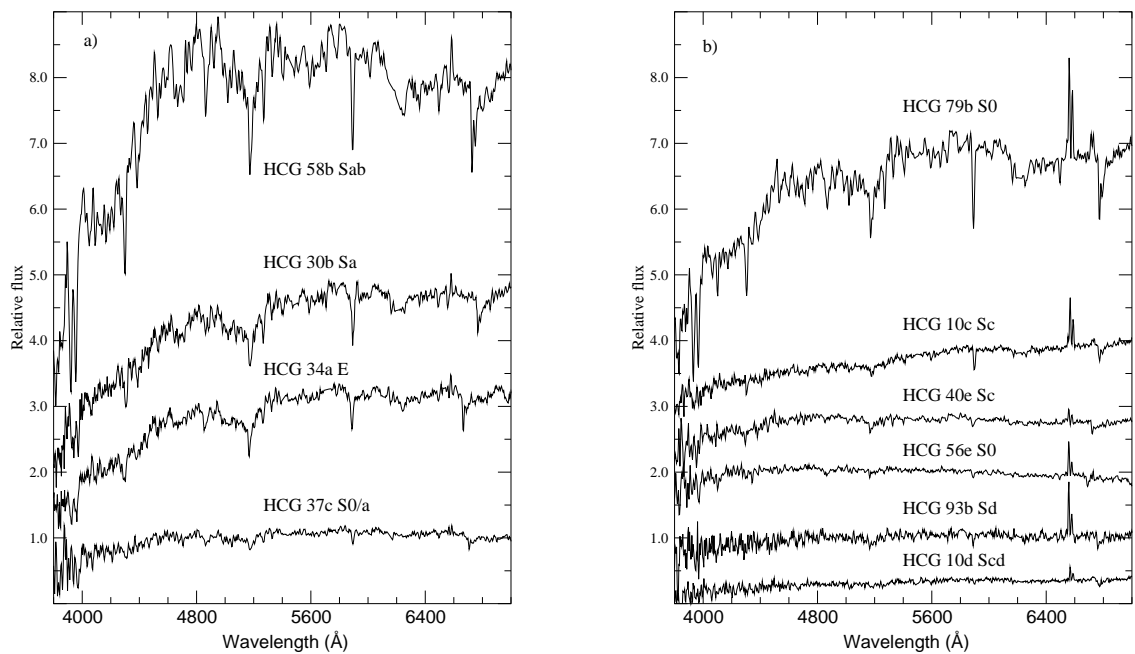}{rcoziol.fig5b.eps}
\caption{a) Spectra of (a) the LLAGN and (b) the SFG candidates. All spectra
are reduced to the restframe, and slightly shifted in flux to facilitate
comparison. The LLAGN candidates share with the confirmed LLAGNs displayed in
Fig.~4 an early--type morphology of the host galaxies. The SFG candidates, with
the exception of HCG 79b and 56e, are all late--type spirals.}
\end{figure}

\clearpage
\begin{figure}
\epsscale{0.80} \figurenum{6}\plotone{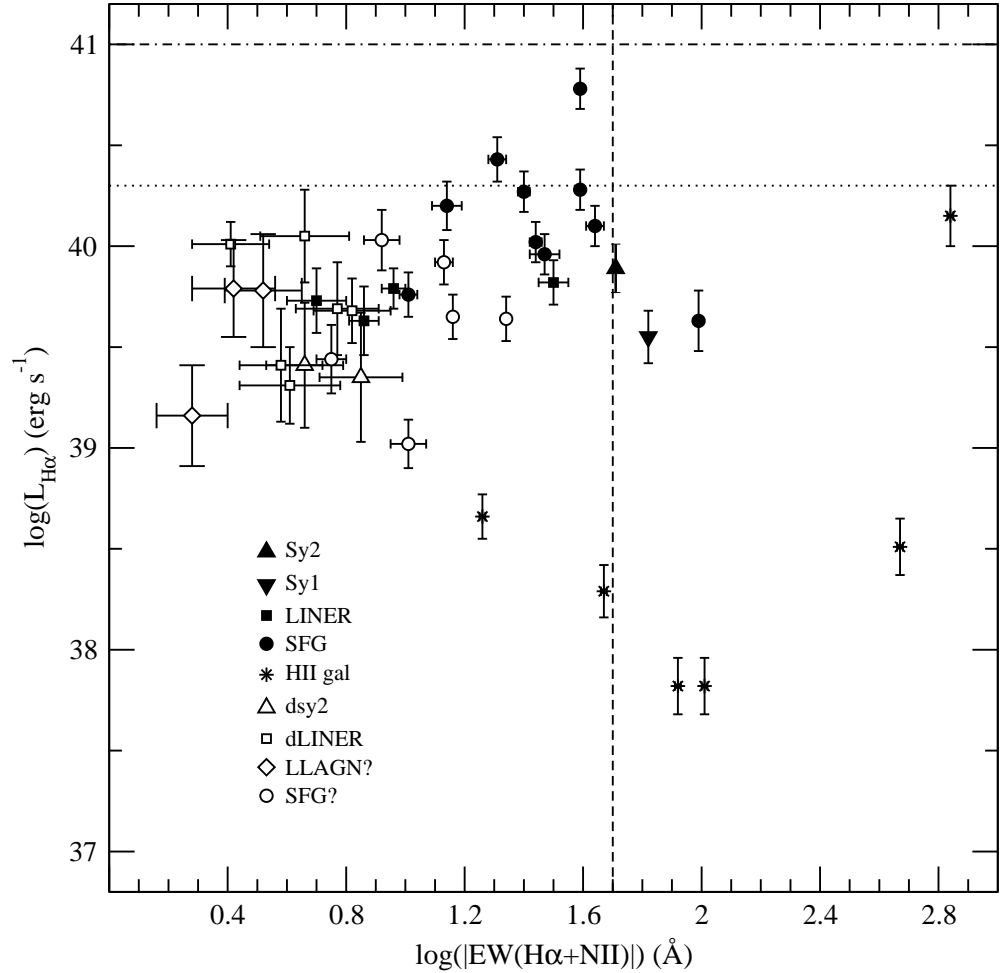} \caption{H$\alpha$
luminosity of the emission line galaxies (after template subtraction) as a
function of the EW of the (H$\alpha+$[NII]) lines. The two horizontal lines are
the mean luminosity for Markarian starburst galaxies (dash-dot) and their lower
limit (dotted). The vertical (dashed) line marks the EW lower limit for
actively star-forming galaxies.}
\end{figure}

\clearpage

\begin{figure}
\epsscale{0.80} \figurenum{7}\plotone{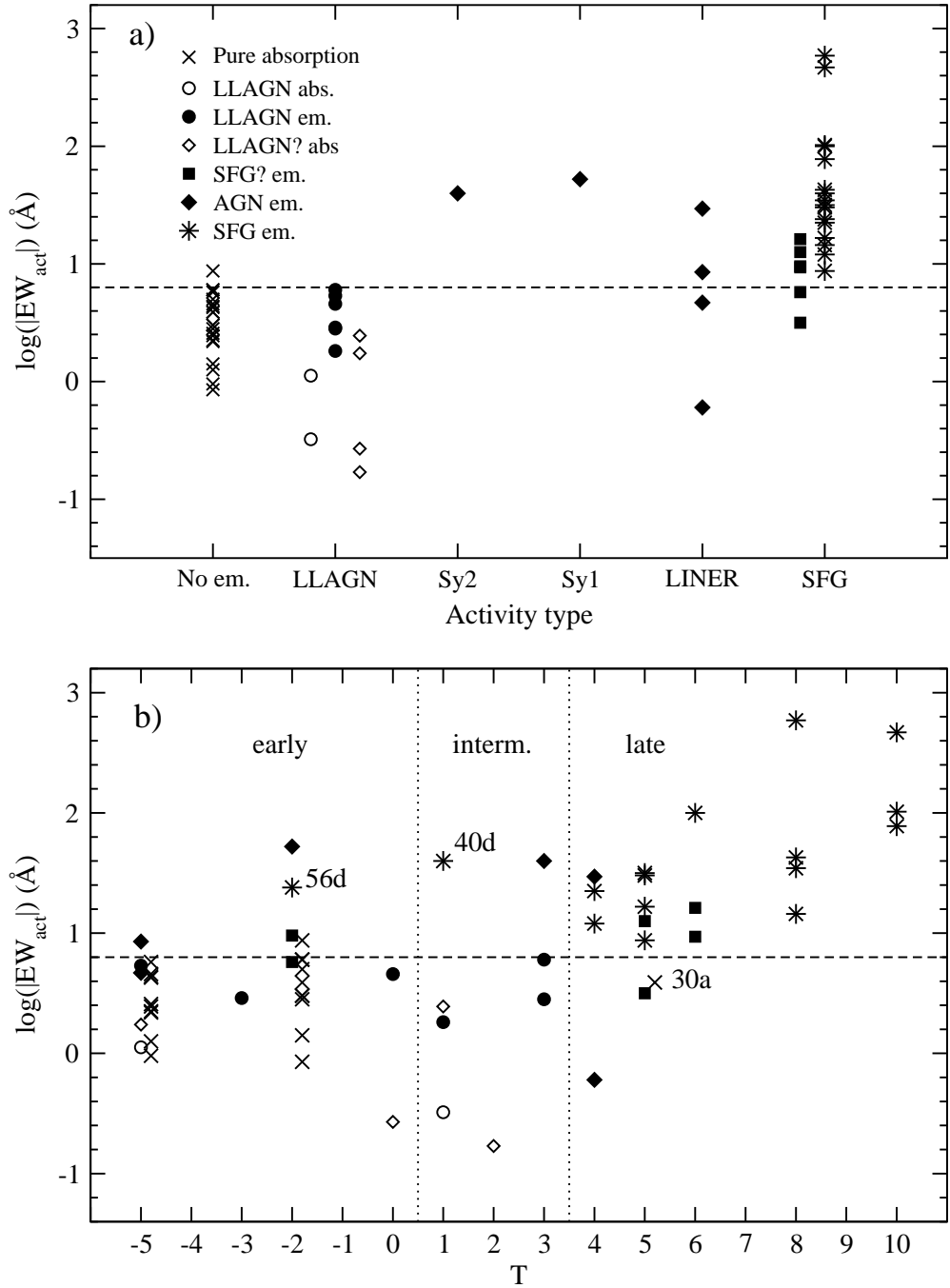} \caption{Value of the
activity index in galaxies with different activity types (a) and morphological
types (b). The horizontal dashed line separates SFGs from quiescent and LLAGNs.
Using different symbols we also distinguish between absorption (EW$_{act}>0$)
and emission (EW$_{act}<0$). The vertical (dotted) lines separate the
morphology axis in three categories: early, intermediate and late-type.}
\end{figure}

\clearpage

\begin{figure}
\epsscale{0.80} \figurenum{8}\plotone{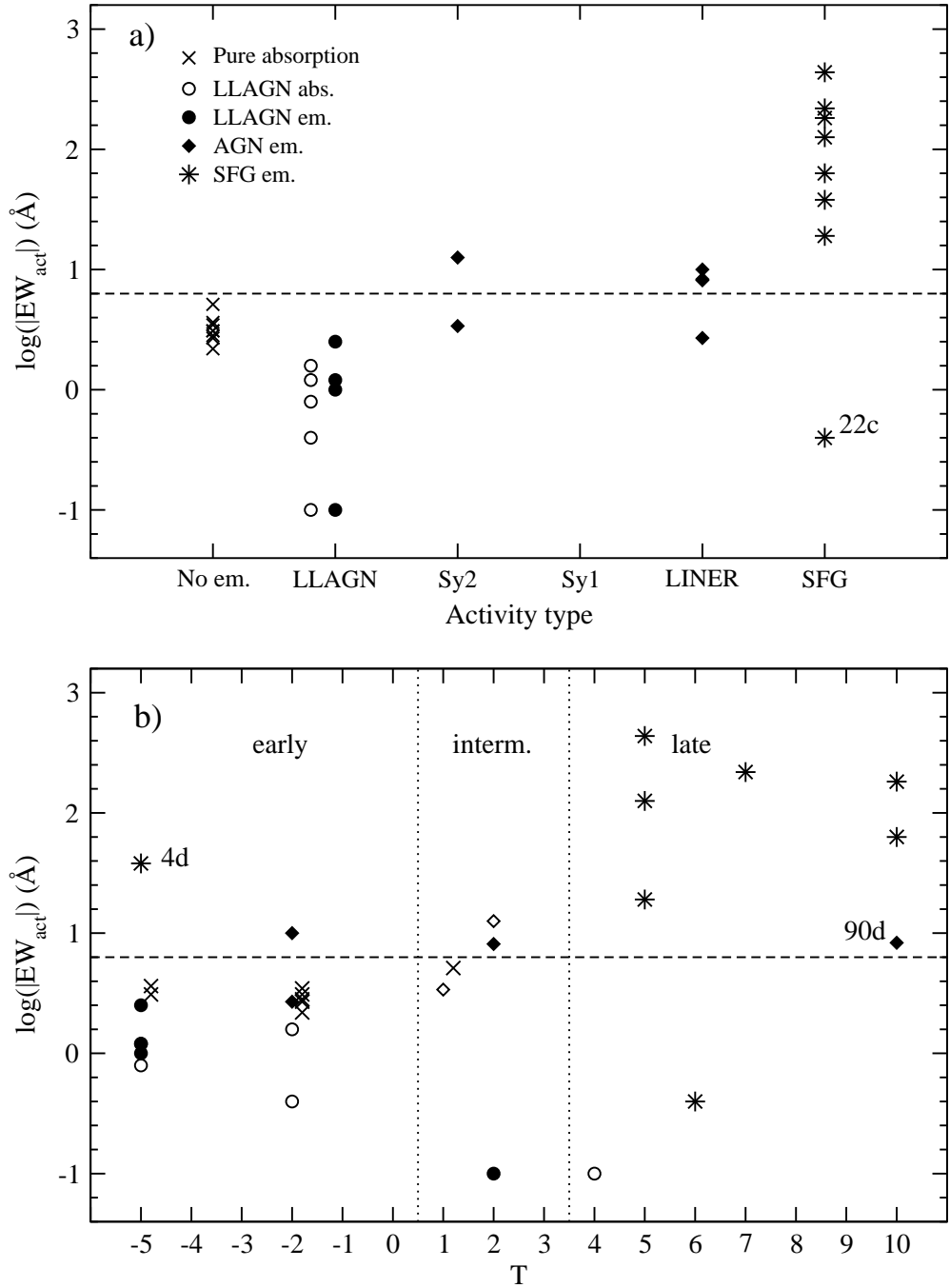} \caption{Value of the
activity index as a function of activity types (a) and morphology (b) for the
galaxies in the additional sample. The meaning of the symbols is the same as in
Figure~7. }
\end{figure}

\clearpage

\begin{figure}
\epsscale{0.8} \figurenum{9}\plotone{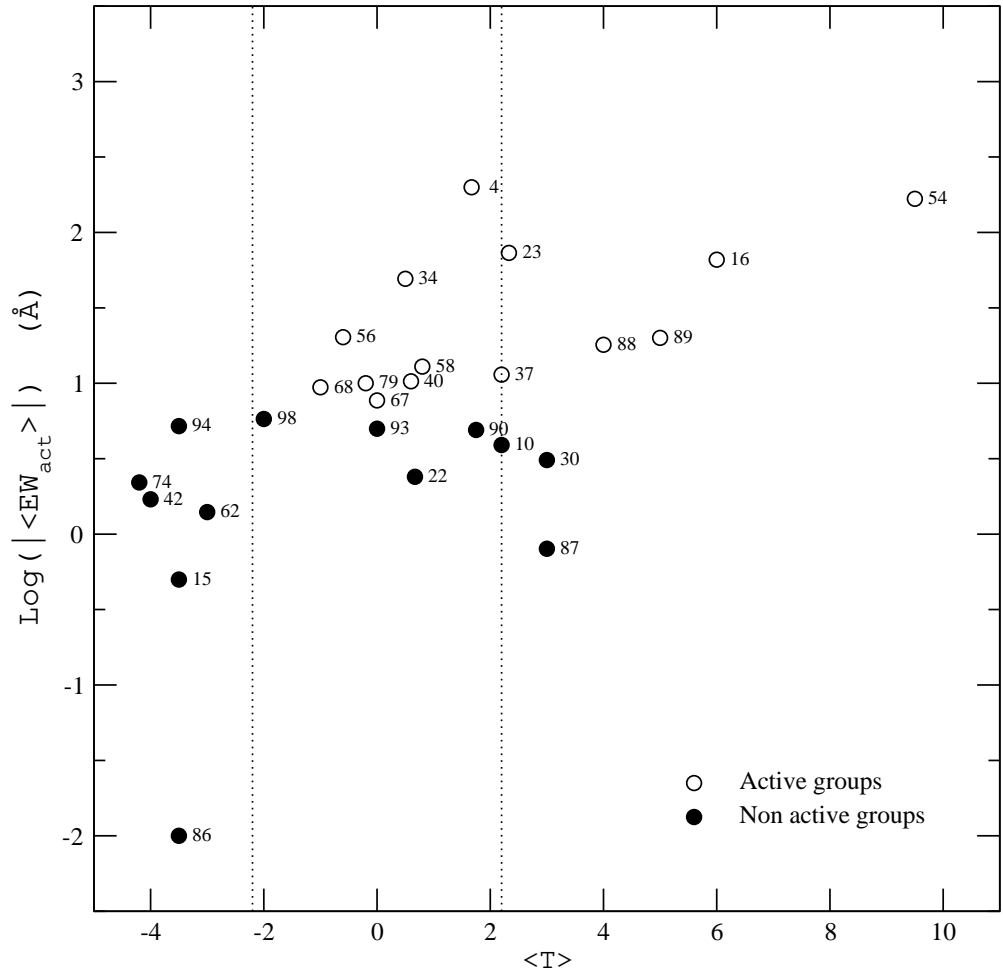} \caption{Mean activity
index as a function of mean morphology. The CGs are identified by their HCG
number. The vertical dot lines separates the groups according to the three main
spatial (projected) configurations. The evolution of groups increases in the
sense A$\Rightarrow$B$\Rightarrow$C}
\end{figure}

\clearpage

\begin{figure}
\epsscale{0.90} \figurenum{10}\plotone{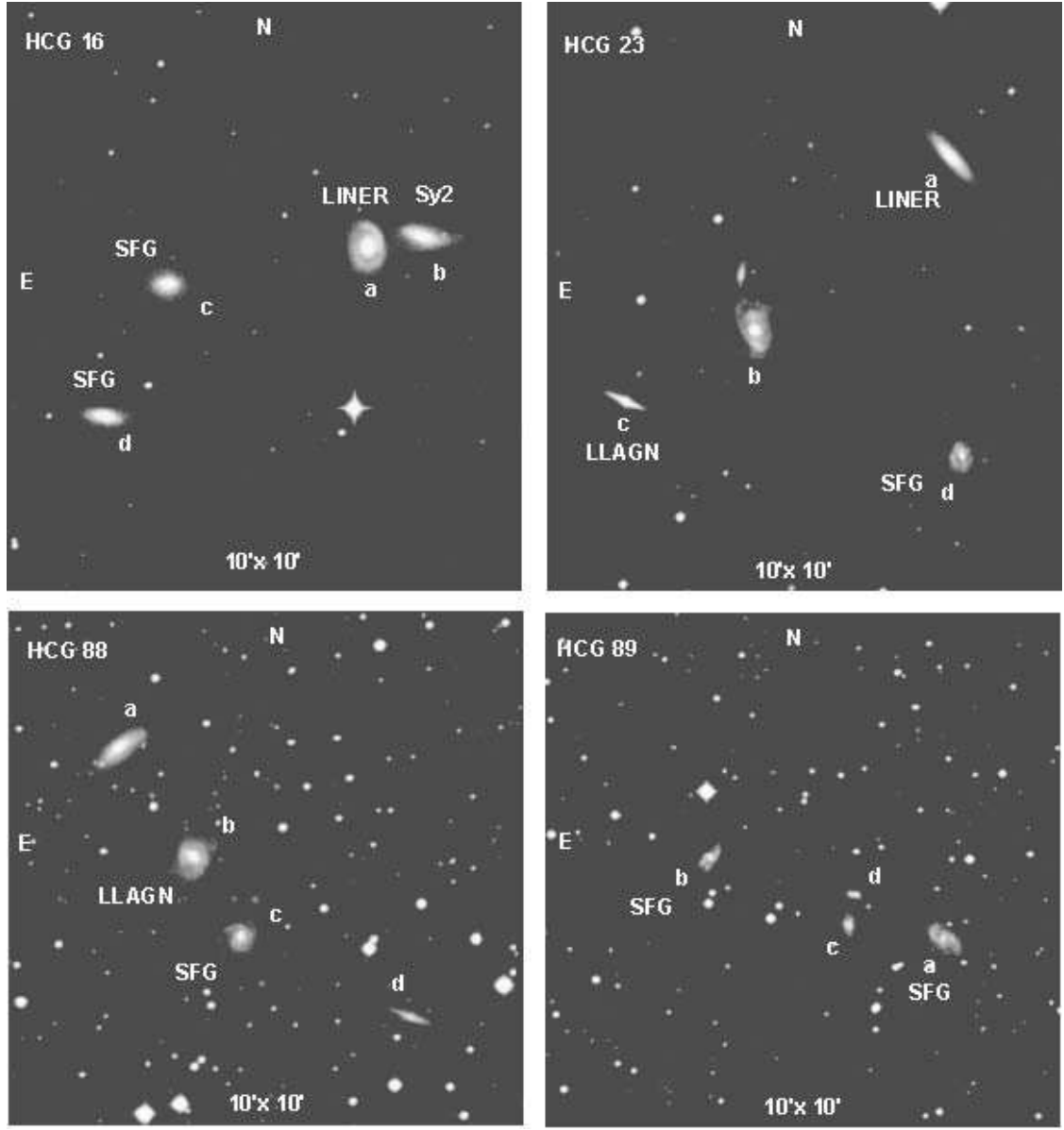} \caption{Type A
prototypes: HCG~16, 23, 88 and 89. The letters identifying individual group
members follow the notation given by Hickson (1982). The activity types from
Table~2 are indicated next to their respective galaxy. The angular size of each
frame is indicated along the bottom.}
\end{figure}

\clearpage

\begin{figure}
\epsscale{0.90} \figurenum{11}\plotone{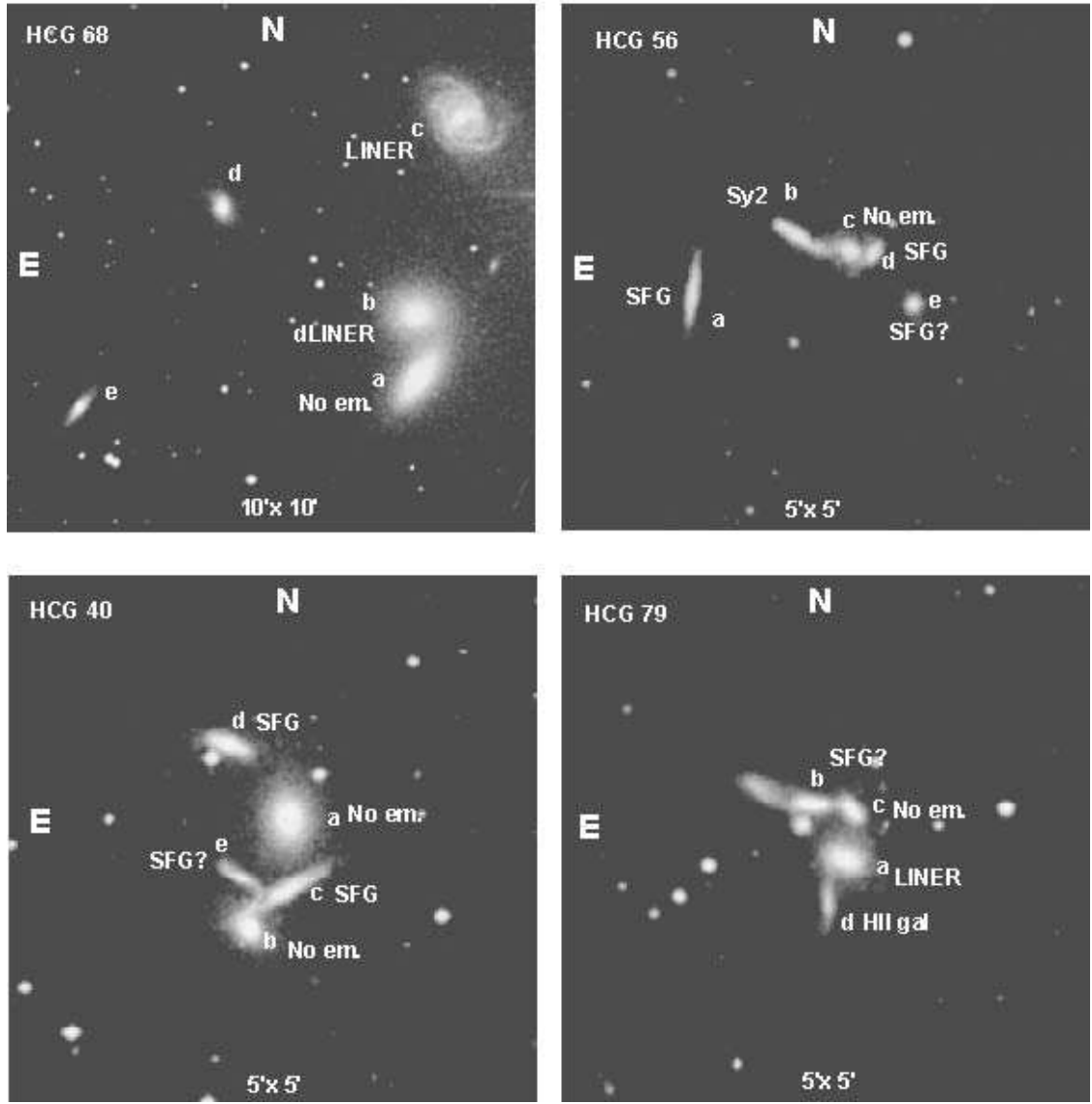} \caption{Type B
prototypes: HCG~40, 56, 68 and 79. Same specifications as in Fig.~10.}
\end{figure}

\clearpage

\begin{figure}
\epsscale{0.90} \figurenum{12}\plotone{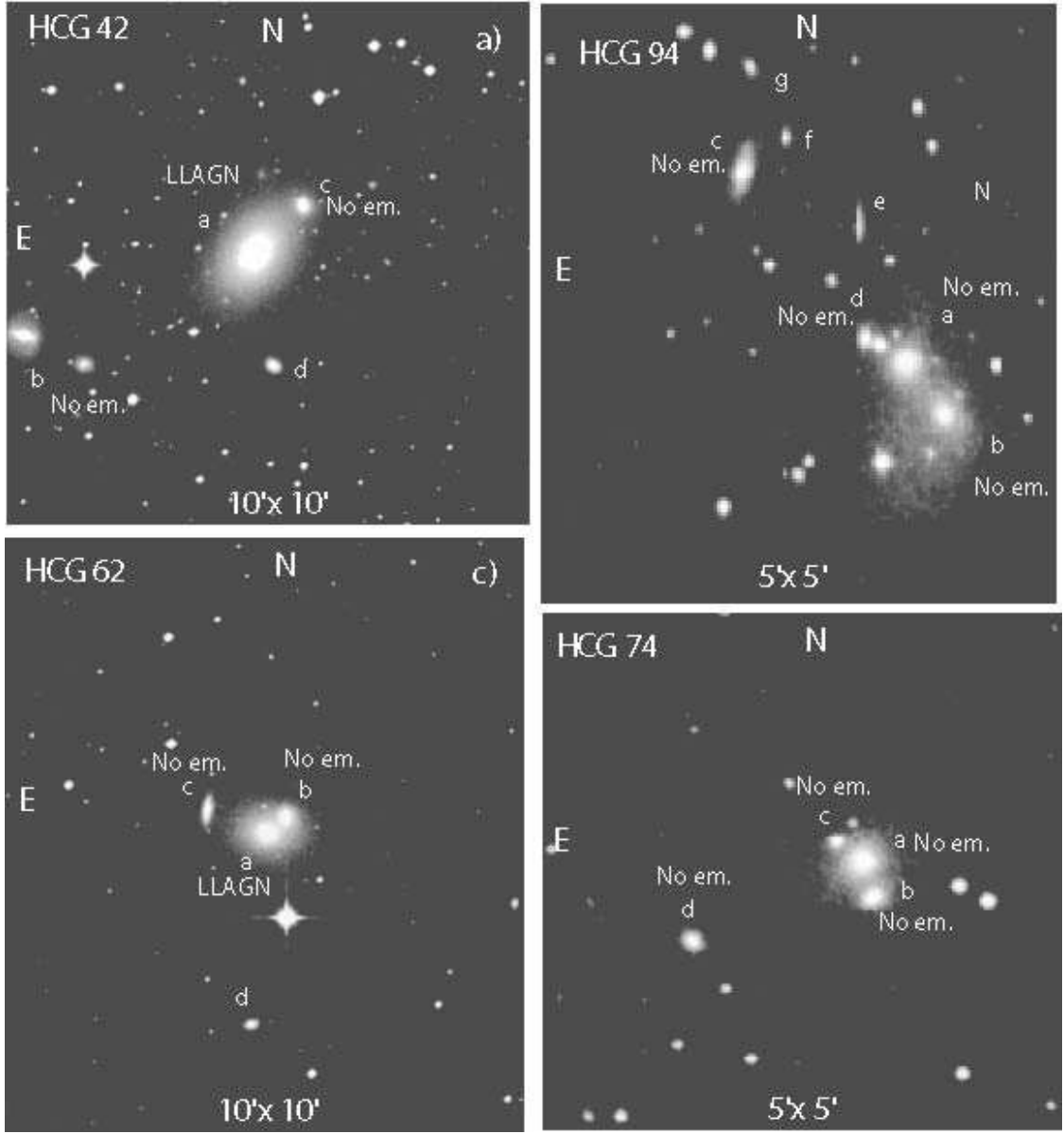} \caption{Type C
prototypes, where one giant elliptical dominates the group: HCG~42, 62, 74 and
94. Same specifications as in Fig.~10.}
\end{figure}

\clearpage

\begin{figure}
\epsscale{0.50} \figurenum{13}\plotone{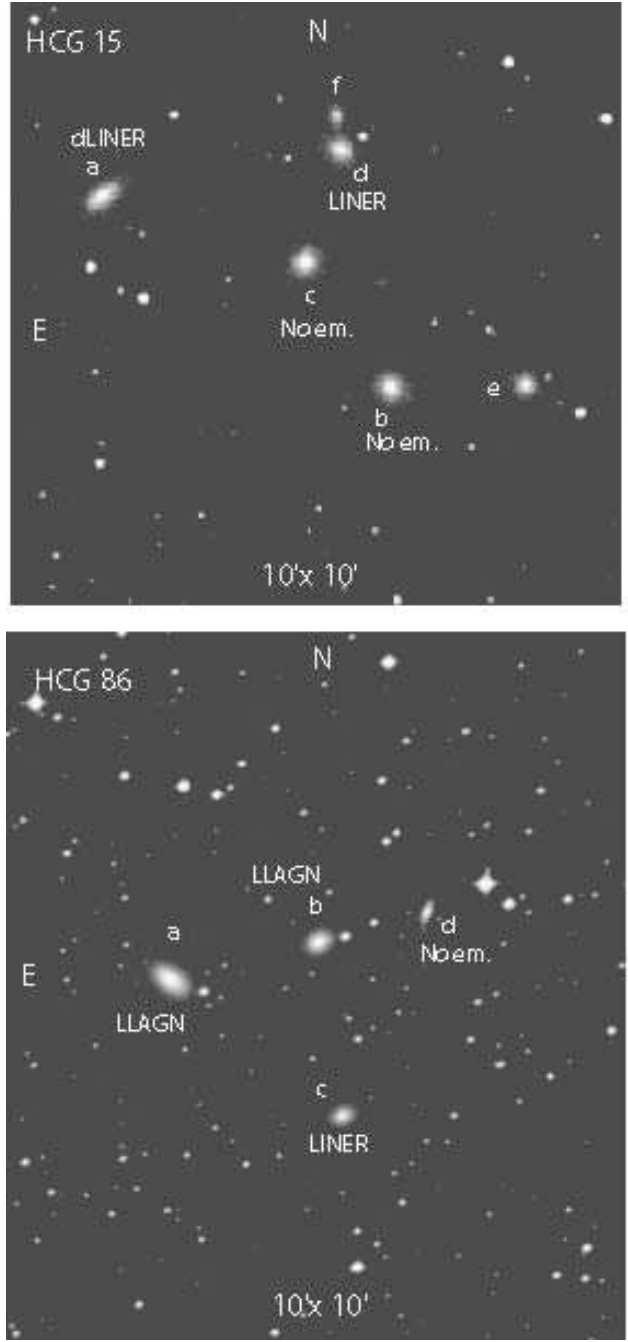} \caption{Type C
prototypes, which are dominated by several elliptical galaxies: HCG~15 and 86.
Same specifications as in Fig.~10.}
\end{figure}

\clearpage

\begin{figure}
\epsscale{0.9} \figurenum{14}\plotone{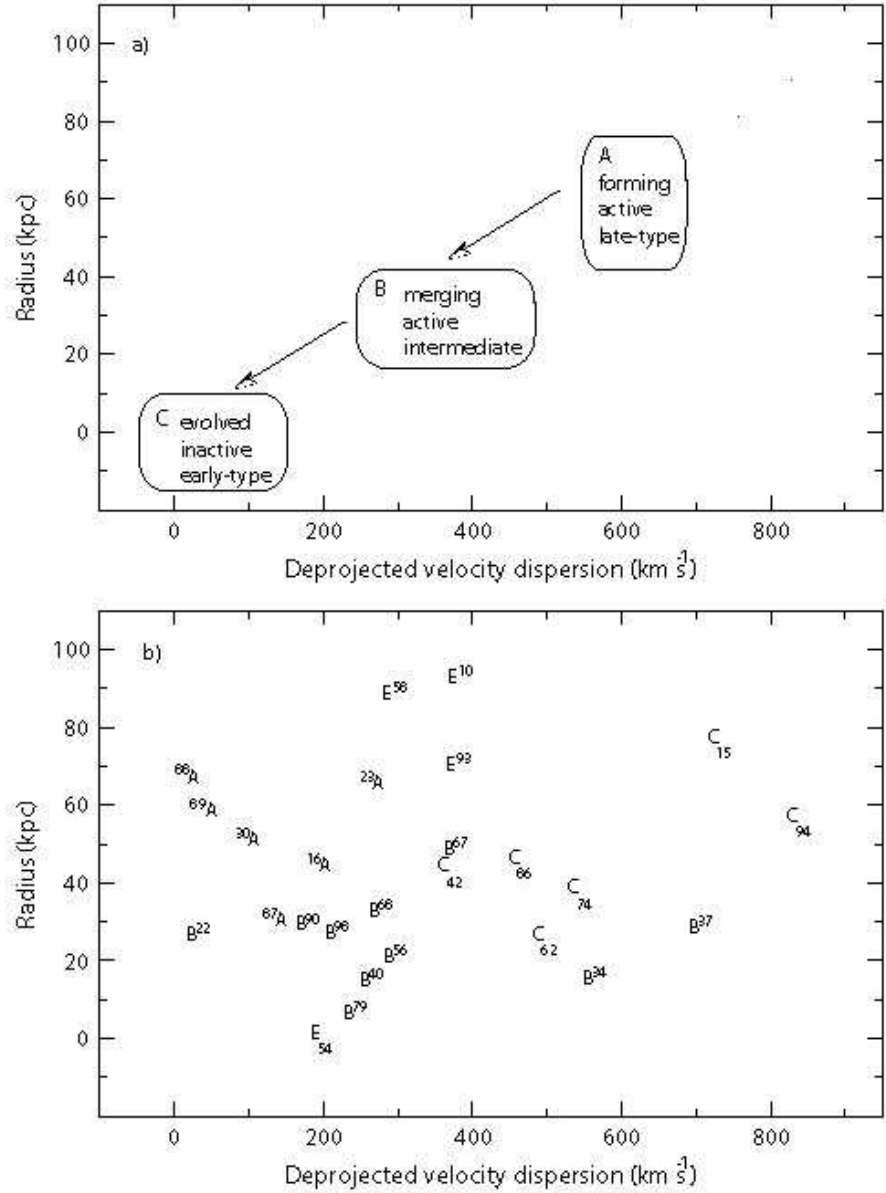} \caption{a)
Illustration of the expected evolutionary pattern, according to the
conventional fast merger model for CGs, in a diagram of radius versus velocity
dispersion; b) actual positions occupied by the 26 CGs (excluding HCG~4) in our
sample. The CGs are identified by their HCG number; the letters correspond to
the configuration types. The E means an exception (no classification is
possible). }
\end{figure}

\clearpage

\begin{figure}
\epsscale{1.0} \figurenum{15}\plotone{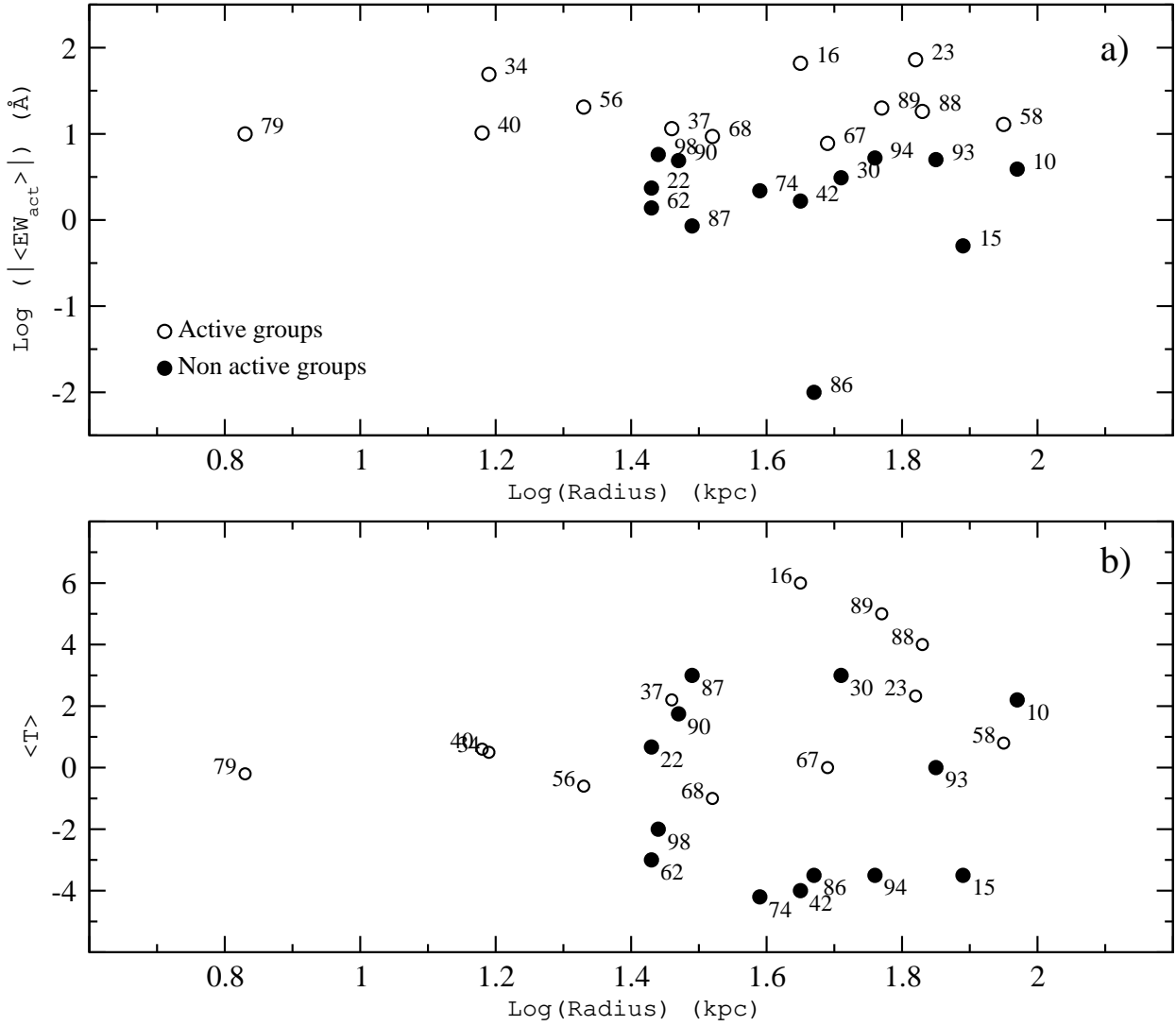} \caption{a) mean
activity index as a function of radius. b) mean morphological type as a
function of radius.}
\end{figure}

\clearpage

\begin{figure}
\epsscale{1.0} \figurenum{16}\plotone{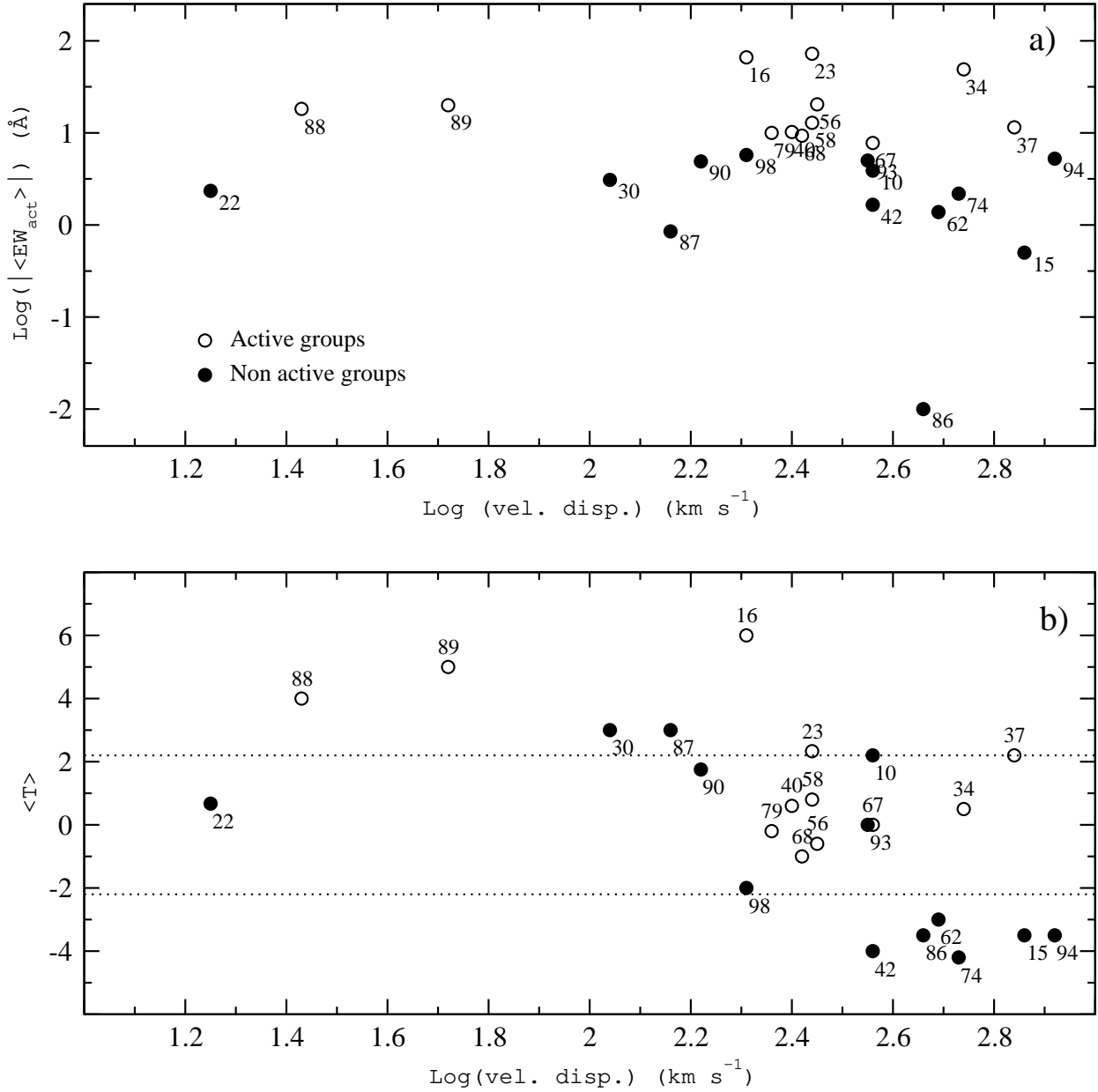} \caption{a) mean
activity index as a function of radius. b) mean morphological type as a
function of radius. The two dotted lines separates the group with different
configuration types and levels of evolution (see Figure~9).}
\end{figure}

\end{document}